\documentclass[pra,twocolumn,a4paper,showpacs]{revtex4}
\usepackage{latexsym}
\usepackage{amsmath}
\usepackage{amsfonts}
\usepackage{graphicx}
\usepackage{exscale}
\usepackage{amssymb}
\usepackage{mathrsfs}
\usepackage{natbib}
\newcommand{\ket}[1]{\ensuremath{|#1\rangle}}
\newcommand{\bra}[1]{\ensuremath{\langle #1 |}}

\newcommand{\ve}{\varepsilon}
\newcommand{\eps}{\epsilon}
\newcommand{\vro}{\varrho}
\newcommand{\be}{\begin{equation}}
\newcommand{\ee}{\end{equation}}
\newcommand{\ba}{\begin{eqnarray}}
\newcommand{\ea}{\end{eqnarray}}
\newcommand{\vfi}{\varphi}
\newcommand{\mc}[1]{\ensuremath{\mathcal{#1}}}
\newcommand{\bc}{\begin{center}}
\newcommand{\ec}{\end{center}}
\newcommand{\bi}{\begin{itemize}}
\newcommand{\ei}{\end{itemize}}

\newcommand{\mf}[1]{\boldsymbol{#1}}
\newcommand{\mfi}[1]{\boldsymbol{#1}}
\newcommand{\Sp}[2]{S_{#1 \,+}^{  (#2)}}
\newcommand{\Sm}[2]{S_{#1 \,-}^{  (#2)}}
\newcommand{\near}{\Omega_N}
\newcommand{\far}{\Omega_F}

\begin{document}

\title{Coherent control in a decoherence-free subspace of a collective
multi-level system}
\author{M. \surname{Kiffner}}
\author{J. \surname{Evers}}
\author{C. H. \surname{Keitel}}
\affiliation{ Max-Planck-Institut f\"ur Kernphysik, 
Saupfercheckweg 1, 69117 Heidelberg, Germany}
\pacs{03.67.Pp, 03.67.Mn, 42.50.Fx}

\begin{abstract}
Decoherence-free subspaces (DFS) in systems of dipole-dipole interacting multi-level atoms are investigated theoretically. 
It is shown that the collective state space of two dipole-dipole interacting four-level 
atoms contains a four-dimensional DFS. 
We describe a  method that allows to populate the 
antisymmetric states of the DFS by means of a laser field, without the 
need of a   field gradient between the two atoms.   
We identify these antisymmetric states as long-lived entangled states. 
Further, we show that any 
single-qubit operation between two states of the DFS can be induced 
by means of a microwave field. 
Typical operation times of these qubit rotations can be 
significantly shorter than for a nuclear spin system. 
\end{abstract}

\maketitle

\section{INTRODUCTION}
The fields of quantum computation and quantum information processing 
have attracted a lot of attention due to their  promising  
applications such as the speedup of classical computations~\cite{chuang:95,ekert:96, nielsen:00}. 
Although the physical implementation of basic quantum information processors 
has been achieved recently~\cite{monroe:02}, the  realization of 
powerful and useable devices is still a challenging and as yet unresolved problem. 
A major difficulty arises from the interaction 
of a quantum system with its environment, which 
leads to decoherence~\cite{divincenzo:95,unruh:95}. 
One possible solution to this problem is provided by the concept of 
decoherence-free subspaces (DFS)~\cite{zanardi:97,lidar:98,lidar:dfs,kempe:01,knill:00,shabani:05}. 
Under certain conditions, a subspace of a 
physical system is decoupled from its environment such that 
the dynamics within this subspace is purely unitary. 
Experimental realizations of   DFS  have been achieved 
with photons~\cite{kwiat:00,zhang:06,altepeter:04,mohseni:03} and in 
nuclear spin systems~\cite{viola:01,wei:05,ollerenshaw:03}.
A decoherence-free quantum memory for one qubit has been 
realized experimentally with two trapped ions~\cite{kielpinski:01,langer:05}. 

The physical implementation of most quantum computation 
and quantum information schemes involves the generation of entanglement 
and the realization of quantum gates.
It has been shown that dipole-dipole interacting systems are 
both a resource for entanglement and  suitable candidates for 
the implementation of gate operations 
between two qubits~\cite{bargatin:00,ficek:02,lukin:00,beige:00,brennen:99,jaksch:00,barenco:95}. 
The creation of entanglement in collective two-atom systems is 
discussed in~\cite{bargatin:00,ficek:02}. 
Several schemes employ the dipole-dipole induced energy shifts of 
collective  states to realize quantum gates, for example, 
in systems of two  atoms~\cite{lukin:00,beige:00,brennen:99,jaksch:00} 
or quantum dots~\cite{barenco:95}. 
In order to ensure that the induced dynamics is fast as compared 
to decoherence processes,  
the dipole-dipole interaction must be strong, and thus 
the distance between the particles must be small. 
On the other hand, it is well known that a system of 
particles which are closer together than the relevant transition wavelength  
displays collective states which are immune against 
spontaneous emission~\cite{agarwal:qst2,ficek:int,mandel:qo,ficek:02,dicke:54}. 
The space spanned by these  subradiant states  is an 
example for a DFS, and hence the question arises whether 
qubits and gate operations enabled by the coherent part of 
the dipole-dipole interaction can be embedded into this DFS. 
In the simple model of a pair of interacting two-level systems,  
there exists only a single subradiant state. 
Larger DFS which are suitable for the storage and 
processing of quantum information can be found, e.g.,  in systems of many 
two-level systems~\cite{zanardi:97b,duan:98b}. 

\begin{figure}[b!]
\includegraphics[scale=1]{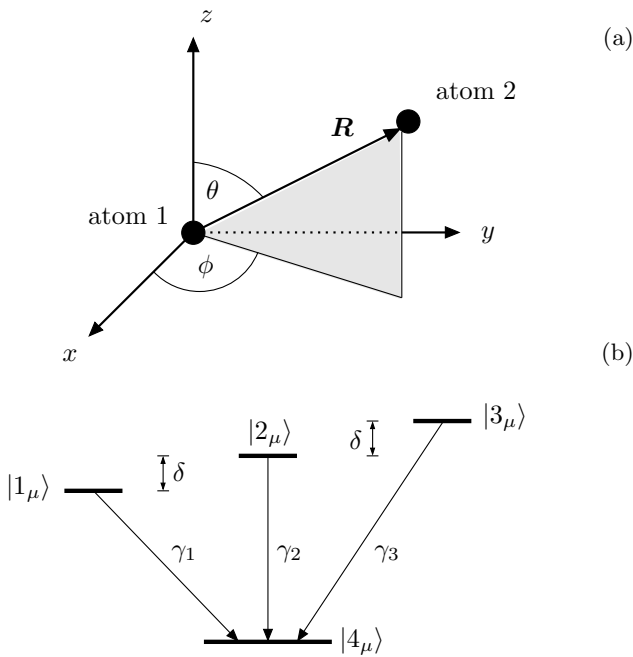}
\caption{\label{picture1} \small   (a) The system under consideration 
is comprised of two atoms that are located at $\mf{r}_1$ and $\mf{r}_2$, respectively. 
The relative position $\mf{R}=\mf{r}_2 -\mf{r}_1$  of atom 2 with respect to atom 1 
is expressed in terms of spherical coordinates.   
(b) Internal level structure of atom $\mu\in\{1,2\}$.  The ground state 
of each of the atoms is a $S_0$ state, and the three excited levels 
are Zeeman sublevels of a $P_1$ triplet. 
The states $\ket{1_{\mu}}$, $\ket{2_{\mu}}$ and $\ket{3_{\mu}}$ 
correspond to the magnetic quantum numbers 
\mbox{$m_j=-1,\,0$} and 1, respectively. 
The frequency splitting of the upper levels is denoted by 
$\delta=\omega_3-\omega_2=\omega_2-\omega_1$, 
where $\hbar \omega_i$ is the energy of state $\ket{i_{\mu}}$. 
  }
\end{figure}

Here, we pursue a different approach and consider a pair of 
dipole-dipole interacting multi-level atoms [see Fig.~\ref{picture1}]. 
The level scheme of each of the atoms is modeled by 
a $S_0\leftrightarrow P_1$ transition 
that can be found, e.g., in ${}^{40}$Ca atoms. 
The excited state multiplet $P_1$ consists of three Zeeman sublevels, and 
the ground state is a $S_0$ singlet state. 
We consider arbitrary geometrical alignments of the atoms, i.e. the length and orientation of 
the vector $\mf{R}$ connecting the atoms can be freely adjusted. 
In this case, all Zeeman sublevels of the atomic multiplets have to 
be taken into account~\cite{kiffner:06}.  
Experimental studies of such systems 
have become feasible  recently~\cite{devoe:96,hettich:02,eschner:01}.  

As our main results, we demonstrate that  the state space 
of the two atoms contains a 4-dimensional DFS  
if the interatomic distance $R$ approaches zero. 
A careful analysis of both the coherent and the incoherent
dynamics reveals that the antisymmetric states of the DFS can be populated with
a laser field, and that coherent dynamics can be
induced within the DFS via an external static magnetic 
or a radio-frequency field. 
Finally, it is shown  that the system can be prepared in 
long-lived entangled states. 

More specifically, all features of the collective two-atom system will be 
derived from the master equation for the two atoms 
which we discuss in Sec.~\ref{EOM}.  
To set the stage, we  prove the existence of 
the 4-dimensional DFS  in the case of small 
interatomic distance $R$  in Sec.~\ref{DFS}. 

Subsequent sections of this paper address the question whether 
this DFS can be employed to store and process quantum information. 
In a first step, we provide a detailed analysis of 
the coherent and incoherent system dynamics  (Sec.~\ref{diagonalization}). 
The eigenstates and energies in the case where the Zeeman splitting 
$\delta$ of the excited states vanishes are presented in Sec.~\ref{degenerate}. 
In Sec.~\ref{section_decay}, we calculate the decay rates of the collective 
two-atom states which are formed by the coherent part of the dipole-dipole 
interaction. It is shown that spontaneous emission in the DFS is strongly suppressed  
if the distance between the atoms  is small 
as compared to the wavelength of the $S_0\leftrightarrow P_1$ transition. 
The full energy spectrum in the presence of a magnetic field 
is investigated in Sec.~\ref{non_degenerate}. 


The DFS is comprised of the collective ground state and 
three antisymmetric collective states. 
In Sec.~\ref{population}, we  
show that the antisymmetric states can 
be populated selectively by means of an external laser field. 
The probability to find the system in a (pure) antisymmetric state 
is 1/4 in steady state.  In particular, the described method 
does not require a field gradient between 
the position of the two atoms. 

We then address coherent control within the DFS,
and demonstrate that the 
coherent time  evolution of two  states in 
the DFS   can be controlled 
via the Zeeman splitting $\delta$ of the excited states and 
therefore by means of an external magnetic field (Sec.~\ref{evolution}). 
Both static magnetic fields and radio-frequency (RF) fields are 
considered.  The time evolution of the two states is visualized in 
the Bloch sphere picture. While a static magnetic field can only induce a 
limited dynamics,  any single-qubit operation can be performed 
by an RF field. 

In Sec.~\ref{entanglement}, we determine the degree 
of entanglement of the symmetric and antisymmetric  collective states 
which are formed by the coherent part of the dipole-dipole interaction. 
We employ the concurrence as a measure of entanglement and show 
that the symmetric and antisymmetric states are 
entangled. The degree of entanglement of the collective states  is the 
same as in the case of two two-level atoms. 
But in contrast to a pair of two-level atoms, 
the symmetric and antisymmetric states of our 
system  are not maximally entangled.  

A brief summary and discussion of our results 
is provided in Sec.~\ref{discussion}.

\section{EQUATION OF MOTION \label{EOM}}
In the absence of laser fields, the system Hamiltonian  is 
given by 
\be
H= H_A +H_F +H_{\text{vac}} \,,
\label{h_total}
\ee
where 
\ba
H_A  & = &  \hbar \sum\limits_{i=1}^3  \sum\limits_{\mu=1}^{2}\omega_i \, \Sp{i}{\mu} \Sm{i}{\mu}\,,
 \nonumber\\[0.2cm]
H_F & = & \sum\limits_{\mfi{k}s}\hbar \omega_k a_{\mfi{k}s}^{\dagger} a_{\mfi{k}s}\,, \nonumber\\[0.2cm]
H_{\text{vac}} & = &  - \mf{\hat{d}}^{(1)}\cdot \mf{\hat{E}}(\mf{r}_{1})- \mf{\hat{d}}^{(2)}\cdot \mf{\hat{E}}(\mf{r}_{2})\, .
\label{H0}
\ea
In these equations, $H_A$ describes the free evolution of the two identical atoms, 
$\hbar\omega_i$ is the energy of state $\ket{i_{\mu}}$  and  
we choose  $\hbar\omega_4=0$.  
The   raising and lowering operators on the \mbox{$\ket{4_{\mu}}\leftrightarrow\ket{i_{\mu}}$} 
transition  of atom $\mu$ are  ($i\in\{1,2,3\}$)
\be
\Sp{i}{\mu} = \ket{i_\mu}\bra{4_{\mu}}  \quad\text{and}\quad  \Sm{i}{\mu} =\ket{4_{\mu}} \bra{i_\mu}\,.
\ee
$H_F$ is the Hamiltonian of the unperturbed vacuum field  and $H_{\text{vac}}$ 
describes the interaction of the atom with the vacuum modes in dipole approximation.
The electric field operator $\mf{\hat{E}}$ is defined as 
\be
\mf{\hat{E}}(\mf{r})=i\sum\limits_{\mfi{k}s}
\sqrt{\frac{\hbar\omega_k}{2\ve_0 V}}\mf{\eps}_{\mfi{k}s}e^{i\mfi{k}\cdot\mfi{r}}  a_{\mfi{k}s} +\text{H.c.}\,,
\ee
where $a_{\mfi{k}s}$ ($a_{\mfi{k}s}^{\dagger}$) are the annihilation (creation) operators that 
correspond to a field mode with wave vector $\mf{k}$, polarization $\mf{\eps}_{\mfi{k}s}$ 
and frequency $\omega_k$, and $V$ denotes the quantization volume. 
We determine the electric-dipole moment operator of atom $\mu$ via the 
Wigner-Eckart theorem~\cite{sakurai:mqm} and arrive at
\be
\mf{\hat{d}}^{(\mu)}=\sum\limits_{i=1}^{3}\big[\mf{d}_{i}  \Sp{i}{\mu} +\text{H.c.}\big]\,,
\ee
where the dipole moments $\mf{d}_i  =   \bra{i}\mf{\hat{d}}\ket{4}$ are given by 
\be
\begin{array}{rlrl}
\mf{d}_1  = &   \mc{D}\,\mf{\eps}^{(+)}\,, &
\mf{d}_2   = &  \mc{D}\,\mf{e}_z \,,\\[0.3cm]
\mf{d}_3   =& - \mc{D}\,\mf{\eps}^{(-)} \,;
&\mf{\eps}^{(\pm)}=&\frac{1}{\sqrt{2}}(\mf{e}_x \pm i \mf{e}_y)\,,
\end{array}
\label{dipoles}
\ee
and  $\mc{D}$ is  the reduced dipole matrix element. 
Note that the dipole moments $\mf{d}_i$ do not depend on the 
index $\mu$ since we assumed that the atoms are identical. 

With the total Hamiltonian $H$ in Eq.~(\ref{h_total}) we derive 
a master equation for the reduced atomic density operator $\vro$. 
An involved  calculation that employs the Born-Markov approximation 
yields~\cite{kiffner:06,evers:06,agarwal:01,agarwal:qst2}
\be
\partial_t \vro  =  -\frac{i}{\hbar} [H_A , \vro] -\frac{i}{\hbar} [H_{\Omega} , \vro] +\mc{L}_{\gamma}\vro \,.
\label{master}
\ee
The coherent evolution of  the atomic states is determined by $H_A+H_{\Omega}$, where 
$H_A$ is defined in Eq.~(\ref{H0}). The Hamiltonian $H_{\Omega}$ 
arises from the vacuum-mediated 
dipole-dipole interaction between the two atoms and is given by 
\ba
\hspace*{-0.5cm} H_{\Omega}  & = &   -\hbar  \sum\limits_{i=1}^3  \left\{ \Omega_{ii} \Sp{i}{2} \Sm{i}{1}  \,+\,\text{H.c.} \right\} 
\nonumber \\
&& - \hbar\left\{\Omega_{21} \left(\Sp{2}{2} \Sm{1}{1} + \Sp{2}{1} \Sm{1}{2}\right) \,+\,\text{H.c.} \right\}  \nonumber \\
& & - \hbar\left\{ \Omega_{31} \left(\Sp{3}{2} \Sm{1}{1} + \Sp{3}{1} \Sm{1}{2}\right) \,+\,\text{H.c.} \right\}  \nonumber \\[0.1cm]
& &  - \hbar\left\{ \Omega_{32} \left(\Sp{3}{2} \Sm{2}{1} + \Sp{3}{1} \Sm{2}{2} \right) \,+\,\text{H.c.} \right\}\,.
\ea
The coefficients $\Omega_{ij}$ cause an energy shift of the 
collective atomic levels (see Sec.~\ref{diagonalization}) and 
are defined as~\cite{kiffner:06,evers:06,agarwal:01}
\be
\Omega_{ij}  = \frac{1}{\hbar} \left[\mf{d}_i^{\text{T}}\; \text{Re}
 (\overset{\leftrightarrow}{\chi})  \; \mf{d}_j^* \right] \,.
\label{Omega}
\ee
Here $\overset{\leftrightarrow}{\chi}$ is a tensor whose components 
$\overset{\leftrightarrow}{\chi}_{kl}$ for $k,l \in\{1,2,3\}$ are given by 
\begin{align}
\overset{\leftrightarrow}{\chi}_{kl}(\mf{R}) & =\frac{k_0^3}{4\pi\ve_0}\left [ \delta_{kl} \left (  
\frac{1}{\eta} + \frac{i  }{\eta^2} - \frac{1}{\eta^3}
\right )\right.    \notag \\[0.2cm]
& \hspace*{1.5cm}\left. -\,\frac{\mf{R}_{k}\mf{R}_{l}}{R^2} \left( \frac{1}{\eta} + \frac{3i}{\eta^2} 
- \frac{3}{\eta^3} \right )\right ]\,e^{i \eta}\,,
\label{chi}
\end{align} 
$\mf{R}$ denotes the  relative coordinates 
of \mbox{atom 2} with respect to \mbox{atom 1}  (see Fig.~\ref{picture1}), and  $\eta=k_0 R$. 
In the derivation of Eq.~(\ref{chi}), the three transition frequencies $\omega_1$, $\omega_2$ and $\omega_3$ have been 
approximated by their mean value $\omega_0=c k_0$ ($c$: speed of light). 
This is justified since the Zeeman splitting $\delta$ is 
small as compared to the resonance frequencies $\omega_i$. 
For $i=j$, the coupling constants in Eq.~(\ref{Omega}) account for the  coherent 
interaction between a dipole of one of the atoms and the corresponding dipole of the other atom. 
Since the 3 dipoles of the system depicted in Fig.~\ref{picture1}(b) are mutually orthogonal [see Eq.~(\ref{dipoles})], 
the terms $\Omega_{ij}$ for $i\not=j$ reflect the interaction between orthogonal 
dipoles of different atoms. The physical origin of these cross-coupling terms 
has been explained in~\cite{evers:06}.

The last term in Eq.~(\ref{master}) accounts for spontaneous emission and reads 
\begin{align}
&\mc{L}_{\gamma} \vro  =           
 -\!\!\sum\limits_{\mu=1}^2\!\sum\limits_{i=1}^3 
\gamma_i \!\left( \Sp{i}{\mu} \Sm{i}{\mu} \vro + \vro \Sp{i}{\mu}\Sm{i}{\mu} - 2 \Sm{i}{\mu} \vro \Sp{i}{\mu}
\right)  \notag \\
&  -\sum\limits_{i=1}^3 \left\{ 
\Gamma_{ii}\left( \Sp{i}{2} \Sm{i}{1} \vro + \vro \Sp{i}{2} \Sm{i}{1} - 2 \Sm{i}{1}\vro \Sp{i}{2}\right) + \text{H.c.}\!\right\}
\notag \\
& - \sum\limits_{\genfrac{}{}{0pt}{2}{\mu,\nu =1}{\mu\not=\nu}}^2\Big\{
\Gamma_{21} \left(
\Sp{2}{\mu} \Sm{1}{\nu} \vro + \vro \Sp{2}{\mu}\Sm{1}{\nu} - 2 \Sm{1}{\nu}\vro \Sp{2}{\mu}\right)
 \notag \\
& \hspace*{1.2cm} +\Gamma_{31} \left(
\Sp{3}{\mu}\Sm{1}{\nu} \vro + \vro \Sp{3}{\mu}\Sm{1}{\nu} - 2 \Sm{1}{\nu}\vro \Sp{3}{\mu}\right) \notag \\[0.2cm]
&\hspace*{1.2cm} + \Gamma_{32} \left(
\Sp{3}{\mu}\Sm{2}{\nu} \vro + \vro \Sp{3}{\mu}\Sm{2}{\nu} - 2 \Sm{2}{\nu}\vro \Sp{3}{\mu}\right)
 \notag\\[0.2cm]
 & \hspace*{2cm} +  \text{H.c.}  \Big\}  \,.
 \label{Lgamma}
\end{align}
The total decay rate of the exited state $\ket{i}$ of each of the atoms is 
given by $2\gamma_i$, where 
\be
\gamma_i = \frac{1}{4\pi\eps_0}\frac{2|\mf{d}_i|^2 \omega_0^3}{3\hbar c^3}=\gamma\,,
\label{gamma}
\ee
and we again employed the approximation \mbox{$\omega_i \approx \omega_0$}.
The collective decay rates $\Gamma_{ij}$ result from the vacuum-mediated dipole-dipole coupling 
between the two atoms and are determined by 
\be
\Gamma_{ij} = \frac{1}{\hbar} \left[\mf{d}_i^{\text{T}} \; \text{Im}
 (\overset{\leftrightarrow}{\chi}) \; \mf{d}_j^* \right]\, .
 \label{Gamma}
 \ee 
The parameters $\Gamma_{ii}$ arise from the interaction 
between a dipole of one of the atoms and the corresponding dipole of the other atom, and 
the cross-decay rates $\Gamma_{ij}$ for $i\not=j$ originate from the 
interaction between orthogonal dipoles of different atoms~\cite{evers:06}.

\begin{figure}[t!]
\bc
\includegraphics[scale=1]{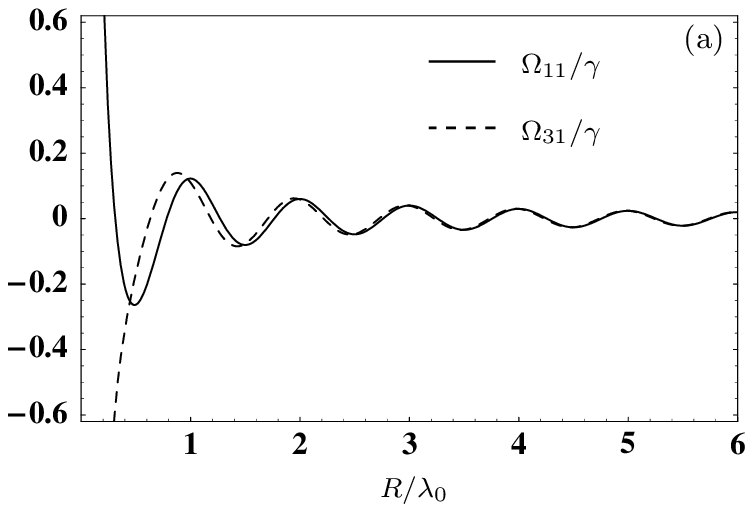}\\[0.3cm]
\includegraphics[scale=1]{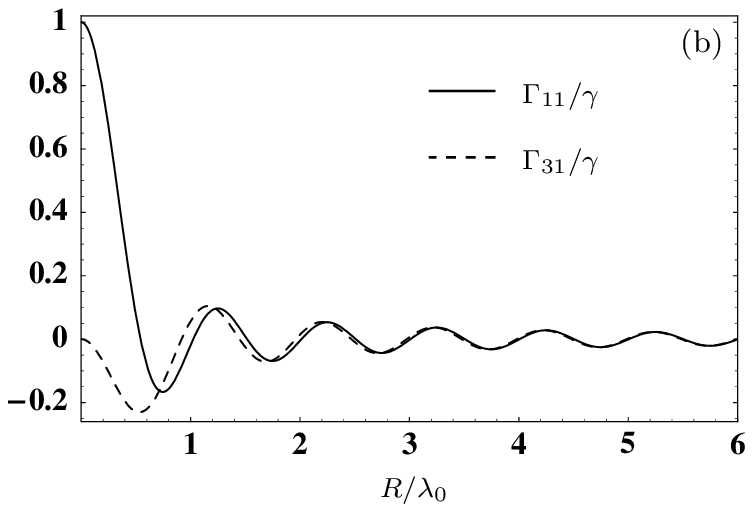}
\caption{\label{picture2} \small  (a) Plot of the vacuum-induced coupling terms  
$\Omega_{11}$ and $\Omega_{31}$  according 
to Eq.~(\ref{OmegaExplicit}). $\lambda_0$ is the mean transition wavelength. If the interatomic 
distance $R$ approaches zero, the parameters $\Omega_{11}$ and $\Omega_{31}$ 
diverge.  (b) Plot of the collective decay rates  
$\Gamma_{11}$ and $\Gamma_{31}$  according to Eq.~(\ref{GammaExplicit}). 
$\Gamma_{11}$ and $\Gamma_{31}$ remain finite in  the limit $R\rightarrow0$. 
The parameters in (a) and (b) are given by $\theta=\pi/2$ and  $\phi=0$.  
  }
\ec
\end{figure}

In order to evaluate the expressions for the various coupling terms $\Omega_{ij}$ and 
the decay rates $\Gamma_{ij}$ in Eqs.~(\ref{Omega}) and (\ref{Gamma}), 
we express the  relative position of the two atoms in spherical coordinates (see Fig.~\ref{picture1}),
\be
\mf{R} =R \, (\sin\theta\cos\phi,\,\sin\theta\sin\phi,\,\cos\theta)\,.
\ee
Together with Eqs.~(\ref{chi}) and ~(\ref{dipoles}) we obtain 
\begin{align}
\Omega_{31} &= \gamma\frac{3}{4\eta^3}\left[\left(\eta^2-3\right)\cos\eta-3\eta\sin\eta\right]\sin^2\theta e^{-2 i \phi}\,,
\notag \\[0.1cm]
\Omega_{11} &  =  3\frac{\gamma}{8\eta^3}\left[\left(3\eta^2-1+\left(\eta^2-3\right)\cos2\theta\right)\cos\eta\right. 
\hspace*{1cm} \notag\\[0.1cm]
& \left.\hspace*{3.5cm} -\eta\left(1+3\cos2\theta\right)\sin\eta\right]\,, \notag \\[0.1cm]
\Omega_{21}    &=     -\sqrt{2} \cot\theta\;\Omega_{31} e^{i\phi}        \,,     \notag \\[0.1cm]
 \Omega_{22}    &=            \Omega_{11} - (2 \cot^2\theta -1) \Omega_{31} e^{2i\phi} \,,  \notag \\[0.2cm]
 \Omega_{32}    &=-\Omega_{21}\,,\quad \Omega_{33}=\Omega_{11}\,,
 \label{OmegaExplicit}
 \end{align}
and the collective decay rates are found to be 
\begin{align}
\Gamma_{31}  & =  \gamma\frac{3}{4\eta^3}\left[\left(\eta^2-3\right)\sin\eta + 3\eta\cos\eta\right]\sin^2\theta e^{-2 i \phi}
\,,\notag \\[0.1cm]
\Gamma_{11} &  =  3\frac{\gamma}{8\eta^3}\left[\left(3\eta^2-1+\left(\eta^2-3\right)\cos2\theta\right)\sin\eta\right. 
\hspace*{1cm} \notag\\[0.1cm]
 &\left.\hspace*{3.5cm} +\eta\left(1+3\cos2\theta\right)\cos\eta\right] \,, \notag\\[0.1cm]
\Gamma_{21} &=      -\sqrt{2} \cot\theta\;\Gamma_{31}e^{i\phi}       \,,       \notag \\[0.1cm]
\Gamma_{22}&=            \Gamma_{11} - (2 \cot^2\theta -1) \Gamma_{31}   e^{2 i\phi} \,,  \notag \\[0.2cm]
 \Gamma_{32} & =-\Gamma_{21}\,,\quad \Gamma_{33}=\Gamma_{11}\,.
 \label{GammaExplicit}
 \end{align}
The coupling terms $\Omega_{11},\,\Omega_{31}$ and the collective 
decay rates $\Gamma_{11},\,\Gamma_{31}$  are shown in 
Fig.~\ref{picture2} as a function of the interatomic distance $R$.

Finally, we consider the case where the two atoms are driven by an external laser field,
\be
\mf{E}_L = \left[\mc{E}_x \mf{e}_x
+ \mc{E}_y \mf{e}_y\right]\,e^{i \mf{k}_L \cdot\mf{r}} \,e^{-i\omega_L t} \,+\,\text{c.c.}\,,
\label{laser_field}
\ee
where $\mc{E}_x$, $\mc{E}_y$ and $\mf{e}_x$, $\mf{e}_y$ denote the field amplitudes 
and polarization vectors, respectively, $\omega_L$ is the laser frequency and c.c. stands for the 
complex conjugate. The  wave vector $\mf{k}_L=k_L\mf{e}_z$ of the laser field  points  in the positive $z$-direction. 
In the presence of the laser field and in a frame rotating with the laser 
frequency, the master equation~(\ref{master})  becomes 
\be
\partial_t \tilde{\vro}  =  -\frac{i}{\hbar} [\tilde{H}_L+\tilde{H}_A , \tilde{\vro}] 
-\frac{i}{\hbar} [H_{\Omega} , \tilde{\vro}] +\mc{L}_{\gamma}\tilde{\vro} \,.
\label{master_L}
\ee
In this equation, $\tilde{H}_A$ is the transformed Hamiltonian of the free atomic evolution, 
\be
\tilde{H}_A  =  - \hbar \sum\limits_{i=1}^3  \sum\limits_{\mu=1}^{2}\Delta_i \, \Sp{i}{\mu} \Sm{i}{\mu}\,.
\label{delta_H}
\ee
The detunings with the state $\ket{i}$ are labeled by $\Delta_i = \omega_L-\omega_i$ ($i\in\{1,2,3\}$), 
and we have $\Delta_1= \Delta_2 +\delta$, $\Delta_3 = \Delta_2-\delta$. 
The Hamiltonian $\tilde{H}_L$  describes 
the atom-laser interaction in the electric-dipole and rotating-wave approximation, 
\ba
\tilde{H}_L  & = & -\hbar \sum\limits_{\mu=1}^{2} \left\{
 \left[\Omega_x(\mf{r}_{\mu}) +i\, \Omega_y(\mf{r}_{\mu})\right] \,\Sp{1}{\mu}    \right.              \nonumber \\[0.2cm]
 && \left.  +   \left[- \Omega_x(\mf{r}_{\mu}) +i\, \Omega_y(\mf{r}_{\mu})\right] \,\Sp{3}{\mu}     \,+\,\text{H.c.} \right\}  \,, 
\ea
and the position-dependent Rabi frequencies are defined as 
\ba
\Omega_x(\mf{r}) & = & \mc{D} \mc{E}_x/(\sqrt{2} \hbar) \,\exp\left[i \mf{k}_L \cdot\mf{r}\right] \,,\nonumber \\[0.2cm]
\Omega_y(\mf{r}) & =  & \mc{D} \mc{E}_y/(\sqrt{2} \hbar) \,\exp\left[i \mf{k}_L \cdot\mf{r}\right]\,.
\ea

\section{DECOHERENCE-FREE SUBSPACE \label{DFS}}
In this section we show that the system depicted in  Fig.~\ref{picture1}  
exhibits a decoherence-free subspace. By definition, 
a subspace  $\mc{V}$ of a Hilbert space $\mc{H}$ is said to 
be decoherence-free if the time evolution inside $\mc{V}$ is 
purely unitary~\cite{lidar:98,lidar:dfs,shabani:05}. 
For the moment, we assume that the system initially is
prepared in a pure or mixed state in the subspace $\mc{V}$. 
The system state is then represented by 
a positive semi-definite Hermitian density operator  
$\vro_{\mc{V}}\in\text{End}(\mc{V})$ 
with $\text{Tr}(\vro_{\mc{V}})=1$.
It follows that $\mc{V}$ is
a decoherence-free subspace if two 
conditions are met. 
First, the time evolution of $\vro_{\mc{V}}$ can only be unitary 
if the decohering dynamics is zero, and therefore we must have 
\be
\mc{L}_{\gamma} \vro_{\mc{V}} = 0
\label{nullspace}
\ee
for all density operators $\vro_{\mc{V}}$ that represent a physical 
system over $\mc{V}$. 
Second, the unitary time evolution governed by $H_A+H_{\Omega}$ 
must not couple states in $\mc{V}$ to  any states outside of $\mc{V}$.  
Consequently, $\mc{V}$ has to be invariant under the action 
of $H_A+H_{\Omega}$, 
\be
\ket{\psi}\in \mc{V}\quad\Longrightarrow\quad (H_A+H_{\Omega})\ket{\psi}\in \mc{V}\,.
\label{invariant}
\ee
Note that since $(H_A+H_{\Omega})$ is Hermitian, this condition
also implies that it cannot couple states outside of $\mc{V}$
to states in $\mc{V}$.

In a first step we seek a solution 
of Eq.~(\ref{nullspace}). To this end we denote the state 
space of the two atoms by $\mc{H}_{\text{sys}}$ and choose the  16  vectors 
$\ket{i, j}=\ket{i_1}\otimes\ket{j_2}$ ($i,\,j\in\{1,2,3,4\}$) 
as a basis of $\mc{H}_{\text{sys}}$. 
The density operator $\vro$ can then be expanded  in 
terms of the  256 operators 
\be
\ket{i, j}\bra{k, l}\,,\qquad i,\,j,\,k,\,l \in\{1,2,3,4\}\,,
\ee
that constitute a basis in the space of all operators acting on $\mc{H}_{\text{sys}}$, 
\be
\vro= \sum\limits_{i,j=1}^4\sum\limits_{k,l=1}^4 \vro_{ij,kl}\ket{i,j}\bra{k,l}\,.
\ee
It follows that $\vro$ can be regarded as 
a vector with 256 components $\vro_{ij,kl}$ and the  linear superoperator 
$\mc{L}_{\gamma}$ is represented  by a   $256\times256$  matrix. 
Equation~(\ref{nullspace}) can thus be transformed into  a homogeneous system of linear equations 
which can be solved by standard methods. 

For a finite distance of the two atoms, the only exact solution of 
Eq.~(\ref{nullspace}) is given by $\ket{4,4}\bra{4,4}$, i.e. 
only the state $\ket{4,4}$ where each of the atoms occupies 
its ground state is immune against spontaneous emission.   
A different situation arises if the interatomic distance $R$ approaches 
zero. In this case, the collective decay rates obey the relations 
\ba
&&\lim\limits_{R\rightarrow 0}\Gamma_{31}=\lim\limits_{R\rightarrow 0}\Gamma_{32}
=\lim\limits_{R\rightarrow 0}\Gamma_{21}=0\nonumber \\[0.2cm]
&&\lim\limits_{R\rightarrow 0}\Gamma_{11}=\lim\limits_{R\rightarrow 0}\Gamma_{22}
=\lim\limits_{R\rightarrow 0}\Gamma_{33}=\gamma\,.
\label{limits}
\ea
In order to characterize the general  solution of Eq.~(\ref{nullspace}) in the limit $R\rightarrow 0$,  
we introduce the three  antisymmetric states 
\be
\ket{a_i}  =  \frac{1}{\sqrt{2}}\big[\,\ket{i,4}-\ket{4,i}\,\big] \,,
\quad i\in\{1,2,3\}\,,
\label{antisym}
\ee 
as well as the 4 dimensional subspace   
\be
\mc{V}=\text{Span}(\ket{4,4},\,\ket{a_1},\,\ket{a_2},\,\ket{a_3})\,.
\ee
The set of operators acting on $\mc{V}$ forms the 16 dimensional 
operator subspace $\text{End}(\mc{V})$. 
We find  that the  solution of Eq.~(\ref{nullspace}) in the limit $R\rightarrow0$ is determined by 
\be
\mc{L}_{\gamma}\, \hat{O} = 0\quad \Longleftrightarrow \quad\hat{O}\in\text{End}(\mc{V})\,.
\ee
In particular, any  positive semi-definite Hermitian operator  $\vro_{\mc{V}}\in\text{End}(\mc{V})$ 
that represents a state over $\mc{V}$ does not decay by spontaneous emission provided that $R\rightarrow0$. 

We now turn to the case of imperfect initialization, i.e.,
the initial state is not entirely contained in the subspace
$\mc{V}$. Then, states outside of $\mc{V}$ spontaneously decay
into the DFS~\cite{shabani:05}. This strictly speaking
disturbs the unitary time evolution inside the DFS, but
does not mean that population leaks out of the DFS.
Also, this perturbing decay into the DFS only occurs on a short timescale 
on the order of $\gamma^{-1}$ at the beginning of the time 
evolution.

These results can be understood as follows. 
In the   Dicke model~\cite{dicke:54,ficek:02} of two nearby 2-level atoms, 
the antisymmetric collective state  is radiatively stable 
if the interatomic distance approaches zero. 
In the system shown in Fig.~\ref{picture1},  
each of the three allowed dipole transitions in one of the atoms 
and the corresponding transition in the other atom form a system that can be thought of 
as   two 2-level atoms. 
This picture is supported by the fact that  the  cross-decay rates originating from the interaction between orthogonal 
dipoles of different atoms vanish as $R$ approaches zero [see Eq.~(\ref{limits})]. 
Consequently, the suppressed decay of one of the antisymmetric states $\ket{a_i}$ is independent 
of the other states. 

In contrast to the cross-decay rates, the coherent dipole-dipole interaction between orthogonal dipoles of 
different atoms is not negligible as $R$ goes to zero. 
It is thus important to verify condition~(\ref{invariant}) that 
requires $\mc{V}$ to be invariant under the action of   $H_A+H_{\Omega}$.
To show that Eq.~(\ref{invariant}) holds, we 
calculate  the matrix representation of $H_{\Omega}$ in the 
subspace $\mc{A}$ spanned by the antisymmetric states 
\mbox{$\{\ket{a_1},\,\ket{a_2},\,\ket{a_3} \}$}, 
\begin{align}
[H_{\Omega}]_{\mc{A}} & = \hbar\,
\left(
\begin{array}{r@{\hspace{0.5cm}}r@{\hspace{0.5cm}}r }
\Omega_{11}       & \Omega_{21}^*            &  \Omega_{31}^*   \\
\Omega_{21}     & \Omega_{22} & \Omega_{32}^*  \\
\Omega_{31} & \Omega_{32} & \Omega_{33}
\end{array}
\right)\,.
\label{H_Omega_A}
\end{align}
Similarly, we introduce the symmetric states 
\be
\ket{s_i}  =  \frac{1}{\sqrt{2}}\big[\,\ket{i,4}+\ket{4,i}\,\big] \,,
\quad i\in\{1,2,3\}\,,
\label{symmetric}
\ee
and the representation of  $H_{\Omega}$ on the subspace $\mc{S}$ spanned by 
the   states \mbox{$\{\ket{s_1},\,\ket{s_2},\,\ket{s_3} \}$}  
is described by 
\begin{align}
[H_{\Omega}]_{\mc{S}} & = - \hbar\,
\left(
\begin{array}{r@{\hspace{0.5cm}}r@{\hspace{0.5cm}}r }
\Omega_{11}      & \Omega_{21}^*            &  \Omega_{31}^*   \\
\Omega_{21}     & \Omega_{22} & \Omega_{32}^*  \\
\Omega_{31} & \Omega_{32} & \Omega_{33} 
\end{array}
\right)\,.
\label{H_Omega_S}
\end{align}
It is found that $H_{\Omega}$ can be written as 
\ba
H_{\Omega} & = & \sum\limits_{i,j=1}^3 \bra{a_i}H_{\Omega}\ket{a_j}\ket{a_i}\bra{a_j} \nonumber\\[0.1cm]
&& +\sum\limits_{i,j=1}^3 \bra{s_i}H_{\Omega}\ket{s_j}\ket{s_i}\bra{s_j} \,,
\label{expansion_HO}
\ea
i.e., all matrix elements $\bra{a_i}H_{\Omega}\ket{s_j}$ between a symmetric and an antisymmetric 
state vanish. This result implies that $H_{\Omega}$ couples the antisymmetric states among themselves, 
but none of them is coupled to a state outside of $\mc{A}$. Moreover, the ground state $\ket{4,4}$ 
is not coupled to any other state by $H_{\Omega}$. 
It follows that the subspace $\mc{V}$ is invariant under the action of $H_{\Omega}$. 

It remains to demonstrate that $\mc{V}$ is invariant under the action of the free  
Hamiltonian $H_A$ in Eq.~(\ref{H0}). 
With the help of the definitions of $\ket{a_i}$ and $\ket{s_i}$ in Eqs.~(\ref{antisym}) and~(\ref{symmetric}),  
it is easy to verify that $H_A$ is diagonal within the subspaces $\mc{A}$ and $\mc{S}$. 
In particular, $H_A$ does not introduce a coupling between the states 
$\ket{a_i}$ and $\ket{s_i}$,
\begin{align}
\bra{s_i}H_A\ket{a_i} & =\frac{1}{2} \left[\bra{i, 4}H_A\ket{i, 4} - \bra{4, i}H_A\ket{4, i}\right] \notag \\[0.2cm]
& = \frac{\hbar}{2} \left( \omega_i \,\bra{i_1}\Sp{i}{1}\Sm{i}{1}\ket{i_1} -  \omega_i \,\bra{i_2}\Sp{i}{2}\Sm{i}{2}\ket{i_2}\right) 
\notag \\[0.2cm]
& =  0\,.
\end{align} 
Note that these matrix elements vanish 
since we assumed that the two atoms are identical, i.e. 
we suppose that the   energy $\hbar\omega_i$ of the internal state $\ket{i_{\mu}}$ does 
not depend on the index $\mu$ which labels the atoms. 

In conclusion, we have shown  that the system of two nearby four-level atoms 
exhibits a four-dimensional  decoherence-free  
subspace $\mc{V} \subset \mc{H}_{\text{sys}}$ if the interatomic distance $R$ approaches zero. 
However, in any real situation the distance between the two atoms remains finite. 
In this case, condition Eq.~(\ref{nullspace}) holds approximately and spontaneous emission in $\mc{V}$ 
is suppressed as long as $R$ is sufficiently small. In Sec.~\ref{section_decay}, we demonstrate 
that the decay rates of states in $\mc{V}$ are smaller than in the 
single-atom case provided that  $R \lesssim 0.43\times\lambda_0$.

\section{SYSTEM DYNAMICS--EIGENVALUES AND DECAY RATES\label{diagonalization}}
The aim of this section is to determine the energies and   decay rates of the 
eigenstates of  the system Hamiltonian $H_A + H_{\Omega}$. 
In a first step (Sec.~\ref{degenerate}), we determine the eigenstates and eigenvalues of $H_{\Omega}$. 
It will turn out that these eigenstates are also eigenstates of $H_A$, 
provided that the Zeeman splitting of the excited states vanishes ($\delta=0$).  
Section~\ref{section_decay} discusses the spontaneous decay rates of the 
eigenstates of $H_{\Omega}$, and   Sec.~\ref{non_degenerate} is concerned with 
the full diagonalization of $H_A + H_{\Omega}$ for $\delta\not=0$.

\subsection{Diagonalization of $H_{\Omega}$\label{degenerate}}
\begin{figure}[t!]
\bc
\includegraphics[scale=1]{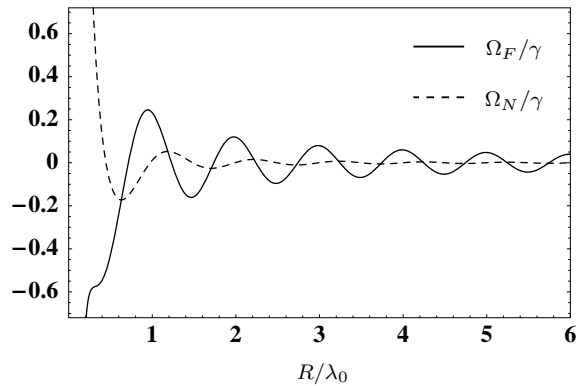}
\caption{\label{picture3} \small  Plot of the vacuum induced energy shifts 
$\far$ and $\near$ 
as a function of the interatomic distance $R$ according to Eq.~(\ref{shifts}). 
These shifts enter the expressions for the 
eigenvalues of $H_{\Omega}$ in Eqs.~(\ref{eigen_a}) and~(\ref{eigen_s}). 
Note that $\far$ decreases with $1/R$ for large values of $R$, while 
$\near$ vanishes with $1/R^2$. 
  }
\ec
\end{figure}

We find the eigenstates and eigenenergies of $H_{\Omega}$ by the 
diagonalization of the two $3\times 3$ matrices 
$ [H_{\Omega}]_{\mc{A}}$ and $ [H_{\Omega}]_{\mc{S}}$ 
which are defined in  Eq.~(\ref{H_Omega_A}) and Eq.~(\ref{H_Omega_S}), 
respectively.  
The eigenstates of 
$H_{\Omega}$ in the subspace $\mc{A}$ spanned 
by the antisymmetric states are given by 
\ba 
\ket{\psi_a^1} 
& = & \sin \theta \ket{a_2}  - \cos \theta \ket{\psi_a^-} \,,\nonumber \\[0.1cm]
\ket{\psi_a^2} 
 & = &  \ket{\psi_a^+}             \,,            \nonumber \\[0.1cm]
\ket{\psi_a^3} 
 & = & \cos \theta \ket{a_2}  +  \sin \theta \ket{\psi_a^-}\,,
 \label{psi_a} 
\ea 
where 
\be 
\ket{\psi_a^{\pm}}   =  \frac{1}{\sqrt{2}}\big[\,e^{i \phi}\ket{a_1} \pm e^{- i \phi}\ket{a_3}\,\big] \,. 
\label{psi_a_pm}
\ee
We denote the  eigenvalue of the state $\ket{\psi_a^i}$  by $\lambda_a^i$ 
and find 
\be
\lambda_a^1   =      \lambda_a^2  
=    \hbar \far  \,,\qquad
\lambda_a^3    =  
   \hbar\near \,,
\label{eigen_a}
\ee
where 
\ba
\far & = & -\gamma \frac{3}{2\eta^3}\left[ \left(1-\eta^2\right)\cos(\eta)+\eta\sin(\eta)\right]\,, \nonumber \\[0.1cm]
\near & = & \gamma  \frac{3}{\eta^3}\left[\cos(\eta)+\eta\sin(\eta)\right]\,,
\label{shifts}
\ea
and $\eta=k_0 R$.  The parameters  $\far$ and $\near$ are shown in Fig.~\ref{picture3} 
as a function of the interatomic distance $R$.  

The eigenstates of  $H_{\Omega}$ in the subspace $\mc{S}$ spanned by the 
symmetric states are found to be 
\ba 
\ket{\psi_s^1} 
& = &  \sin \theta \ket{s_2}  - \cos \theta \ket{\psi_s^-}    \,, \nonumber \\[0.1cm]
\ket{\psi_s^2}
 & = &             \ket{\psi_s^+}   \,,     \nonumber \\[0.1cm]
\ket{\psi_s^3} 
 & = &     \cos \theta \ket{s_2}  + \sin \theta \ket{\psi_s^-}\,,
  \label{psi_s} 
\ea 
where 
\be 
\ket{\psi_s^{\pm}}  =  \frac{1}{\sqrt{2}}\big[\,e^{i \phi}\ket{s_1} \pm e^{- i \phi}\ket{s_3}\,\big] \,,
\label{psi_s_pm}
\ee
and the corresponding eigenvalues read 
\be
\lambda_s^1  = \lambda_s^2  = -      \hbar \far\, , \quad
\lambda_s^3  = -      \hbar \near\,.
\label{eigen_s}
\ee
Next we discuss several features of the eigenstates and eigenenergies of 
$H_{\Omega}$. First, note that two of the symmetric (antisymmetric) states  are 
degenerate.  Second, we point out that 
the  matrices  $ [H_{\Omega}]_{\mc{A}}$ and $ [H_{\Omega}]_{\mc{S}}$  
consist of  the coupling terms $\Omega_{ij}$ which  depend on 
the  interatomic distance $R$ and 
the angles $\theta$ and $\phi$ [see Fig.~\ref{picture1} and Eq.~(\ref{OmegaExplicit})]. 
On the contrary, the eigenstates $\ket{\psi_a^i}$ and $\ket{\psi_s^i}$ depend 
only on the angles $\theta$ and $\phi$, but not on the interatomic 
distance $R$. Conversely, the eigenvalues of $H_{\Omega}$ are only 
functions of the atomic separation $R$ and  do not depend on the 
angles $\theta$ and $\phi$.  
This remarkable result is consistent with a general theorem~\cite{kiffner:06} 
that has been derived for two dipole-dipole interacting atoms. 
The theorem states  that the dipole-dipole induced 
energy shifts between collective two-atom states depend on the 
length of the vector connecting the atoms, but not on its orientation, 
provided that the level scheme of  each atom is modelled by  
complete sets of angular momentum multiplets. 
Since we take all magnetic sublevels of the 
$S_0\leftrightarrow P_1$ transition into account, the theorem 
applies to the system shown in Fig.~\ref{picture1}.

In Sec.~\ref{non_degenerate}, we  show that the eigenstates $\ket{\psi_a^i}$ and 
$\ket{\psi_s^i}$ of $H_{\Omega}$ are also eigenstates of  $H_A$, provided that 
the Zeeman splitting $\delta$ of the excited states vanishes. 
This implies that the energy levels of the degenerate system ($\delta=0$) do not 
depend on the angles $\theta$ and $\phi$, but only on the interatomic distance $R$. 
From a physical point of view, this result can be understood as follows. 
In the absence of a magnetic field ($\delta=0$), there is no distinguished direction in space. 
Since the vacuum is isotropic in free space, one expects that the energy levels 
of the system are invariant under rotations of the separation vector $\mf{R}$.

\begin{figure}[b!]
\bc
\includegraphics[scale=1]{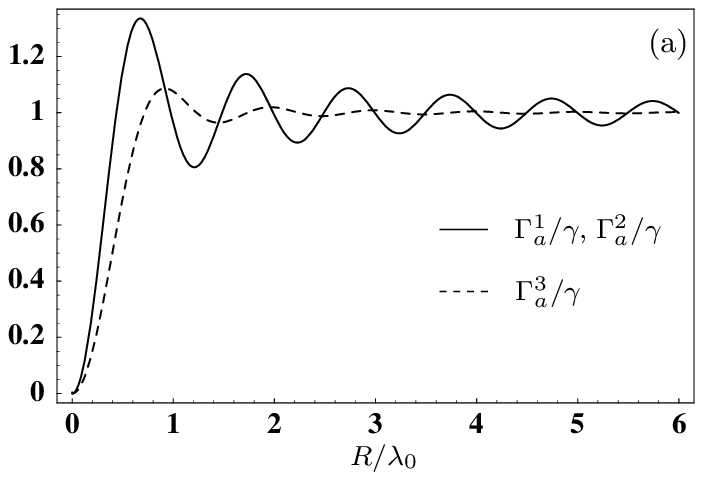}\\[0.3cm]
\includegraphics[scale=1]{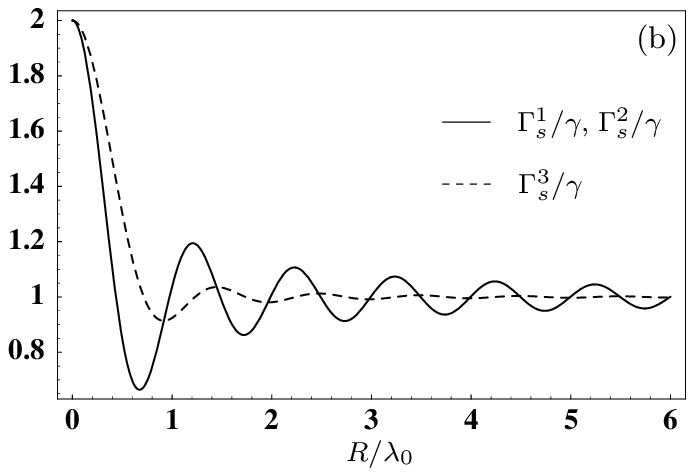}
\caption{\label{picture4} \small 
Dependence of the parameters  $\Gamma_a^i$ and $\Gamma_s^i$ 
on the interatomic distance $R$ according to Eq.~(\ref{decay_rates}). 
(a)  In the limit $R\rightarrow 0$, the $\Gamma_a^i$ tend to zero, 
and the antisymmetric states $\ket{\psi_a^i}$  are subradiant. 
(b) The symmetric states $\ket{\psi_s^i}$ decay twice as fast as 
compared to two independent atoms if  $R$ approaches zero. 
  }
\ec
\end{figure}

\subsection{Decay rates \label{section_decay}}
In order to find the  decay rates that correspond to  the   Eigenstates 
$\ket{\psi_a^i}$ and $\ket{\psi_s^i}$ 
of the  Hamiltonian $H_{\Omega}$, 
we project 
Eq.~(\ref{Lgamma}) onto these states and arrive at 
\ba
\partial_t\,\bra{\psi_a^i}\vro\ket{\psi_a^i} & =&   -2 \,\Gamma_a^i\,\bra{\psi_a^i}\vro\ket{\psi_a^i} + C_a^i(t) \,,
\nonumber  \\[0.1cm]
\partial_t\,\bra{\psi_s^i}\vro\ket{\psi_s^i} & =&   -2\, \Gamma_s^i\,\bra{\psi_s^i}\vro\ket{\psi_s^i} + C_s^i(t)\,.
\label{eom_anti}
\ea
In these equations, $2 \,\Gamma_a^i$ and  $2 \,\Gamma_s^i$  denote the decay rates  
of the  states  $\ket{\psi_a^i}$ and $\ket{\psi_s^i}$, respectively. 
The time-dependent functions $C_a^i(t)$ and $C_s^i(t)$ describe the increase of the  populations 
$\bra{\psi_a^i}\vro\ket{\psi_a^i}$ and $\bra{\psi_s^i}\vro\ket{\psi_s^i}$ due to spontaneous 
emission from  states $\ket{i,j}$  \mbox{($i,j\in\{1,2,3\}$)} where both atoms occupy an excited state. 
The explicit expressions for the  coefficients $\Gamma_a^i$ and $\Gamma_s^i$ as a function 
of the parameter  $\eta=k_0 R$ are given by 
\ba
\Gamma_a^1 &=  & \Gamma_a^2  
   =   \gamma \frac{1}{2\eta^3}\left[2\eta^3  -3 \eta\cos(\eta)+ 3 \left(1-\eta^2\right)\sin(\eta)\right]\,, \nonumber \\[0.1cm]
\Gamma_a^3 & = &  \gamma  \frac{1}{\eta^3}\left[\eta^3 + 3\eta \cos(\eta)- 3\sin(\eta)\right]\,,\nonumber \\[0.1cm]
\Gamma_s^1 &= & \Gamma_s^2  
   =   \gamma \frac{1}{2\eta^3}\left[2\eta^3  + 3 \eta\cos(\eta)- 3 \left(1-\eta^2\right)\sin(\eta)\right] \,,\nonumber \\[0.1cm]
\Gamma_s^3 & = &  \gamma  \frac{1}{\eta^3}\left[\eta^3 -  3\eta \cos(\eta) + 3\sin(\eta)\right]\,.
\label{decay_rates}
\ea
These functions do not depend on the 
angles $\theta$ and $\phi$, but only on the interatomic distance $R$. As for  
the dipole-dipole induced energy shifts 
of the states $\ket{\psi_a^i}$ and $\ket{\psi_s^i}$  (see Sec.~\ref{degenerate}), 
this result is in agreement with the theorem derived in~\cite{kiffner:06}. 

\begin{figure}[t!]
\bc
\includegraphics[scale=1]{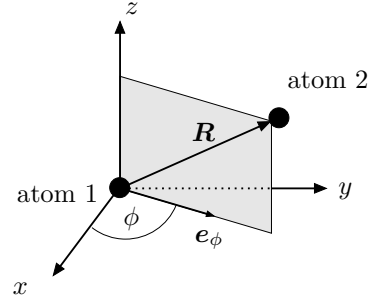}
\caption{\label{picture5} \small 
The atoms are aligned in a plane spanned by 
the unit vectors $\mf{e}_z$ 
and $\mf{e}_{\phi}=(\cos\phi,\sin\phi,0)$. 
Within this plane, the relative position of the two atoms  $\mf{R}= z\,\mf{e}_z + l\, \mf{e}_{\phi}$ is 
described by the parameters $z$ and $l$. The energies of the eigenstates of $H_A + H_{\Omega}$ 
depend only on $z$ and $l$, but not on $\phi$. 
  }
\ec
\end{figure}
\begin{figure*}[t!]
\bc
\hspace*{0.5cm}
\includegraphics[scale=1]{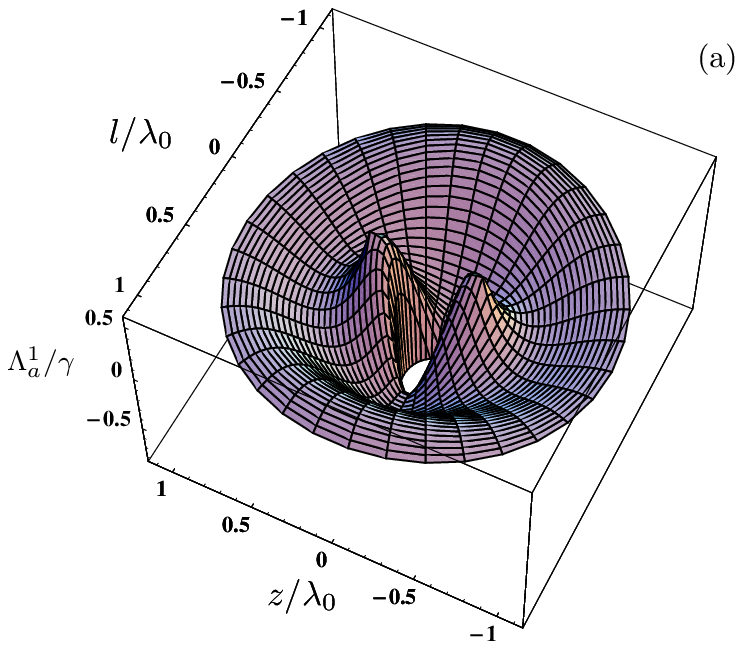}\hspace*{\fill}
\includegraphics[scale=1]{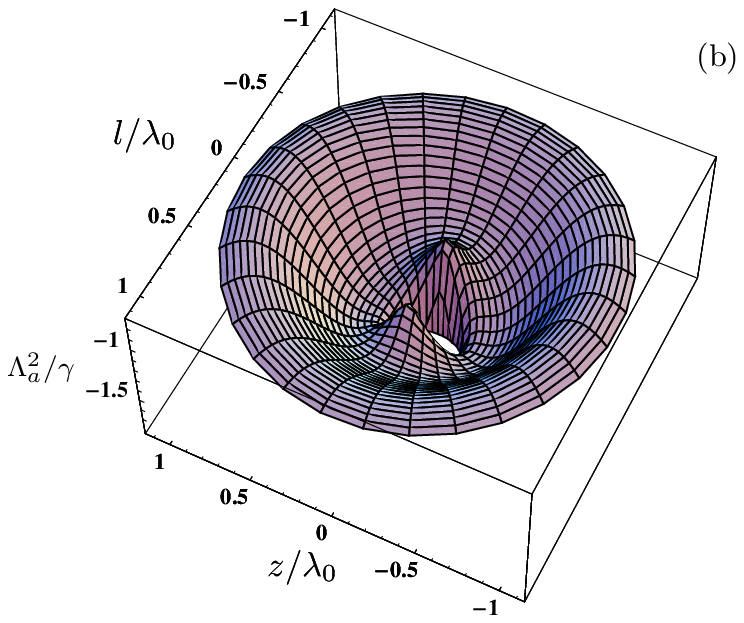} \hspace*{0.7cm} \\[0.3cm]
\hspace*{0.5cm}\includegraphics[scale=1]{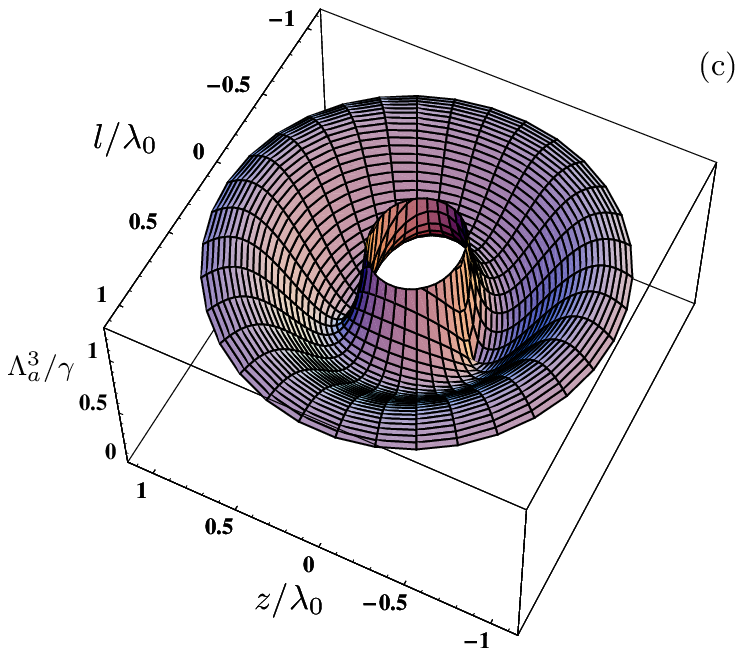}\hspace*{\fill}
\includegraphics[scale=1]{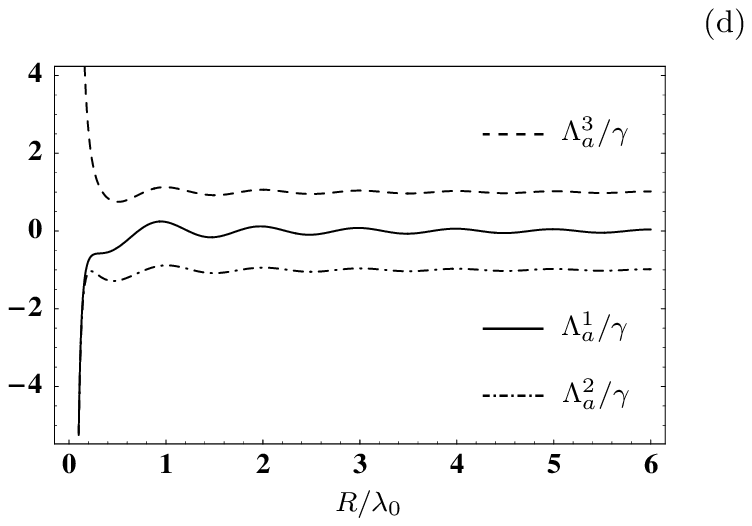}\hspace*{0.7cm}
\caption{\label{picture6} \small   (Color online) Plot of the energy shifts that determine the 
energy levels of the antisymmetric states according to Eq.~(\ref{energy_a}). 
In (a)-(c), the parameters $\Lambda_a^i$ are shown in a plane spanned by $\mf{e}_z$  and 
$\mf{e}_{\phi}=(\cos\phi,\sin\phi,0)$.  The relative position $\mf{R}= z\,\mf{e}_z + l\, \mf{e}_{\phi}$ 
of the atoms in this plane is parameterized by $z$ and $l$ (see also Fig.~\ref{picture5}).  
Since the $\Lambda_a^i$ do not depend on $\phi$, 
the energy surfaces shown in (a)-(c) do not change if  $\mf{e}_{\phi}$ is rotated around the $z$-axis. 
While   $\Lambda_a^1$ and $\Lambda_a^2$  tend 
to $-\infty$ in the limit $R\rightarrow 0$, $\Lambda_a^3$ tends to $+\infty$. 
The frequency splitting of the excited states is $\delta = \gamma$. 
In (d), the $\Lambda_a^i$ are shown as a function of the  interatomic 
distance $R$, the parameters are  $\theta=\pi/2$ and  $\delta = \gamma$.  
}
\ec
\end{figure*}
Figure~\ref{picture4}(a) shows the parameters $\Gamma_a^i$  as a function of $R$. 
The oscillations of 
$\Gamma_a^1$ and $\Gamma_a^2$  around $\gamma$ are damped with $1/R$ as 
$R$ increases,   and  those of $\Gamma_a^3$ decrease with $1/R^2$. 
Note that the oscillations of the  frequency shifts $\lambda_a^i$ 
display similar features for $R\gg\lambda_0$ (see Sec.~\ref{degenerate}).  
It has been shown in Sec.~\ref{DFS} that any state within the 
subspace $\mc{A}$ of antisymmetric states is  completely stable for $R\rightarrow 0$. 
Consequently, the decay rates  $2\Gamma_a^i$ of the states $\ket{\psi_a^i}$ 
tend to zero  as $R$ approaches zero. 
It can be verified by numerical methods that $\Gamma_a^1$ and $\Gamma_a^2$  are smaller 
than the parameter $\gamma$ provided that  $R \lesssim 0.44\times\lambda_0$, and $\Gamma_a^3$ does 
not exceed $\gamma$  if $R \lesssim 0.72\times\lambda_0$. For $R=0.1\times\lambda_0$, the 
coefficients $\Gamma_a^i$ are smaller than $0.1\times \gamma$. 
Although $R$ is larger than zero in an experiment, the states $\ket{\psi_a^i}$ 
decay much slower as compared to two non-interacting atoms  if $R$ is sufficiently small. 
This shows that spontaneous emission can be strongly suppressed  within the subspace $\mc{A}$  
of the  antisymmetric states, even for a realistic value of the interatomic distance $R$.

The parameters $\Gamma_s^i$  are depicted in 
Fig.~\ref{picture4}(b).  In the limit $R\rightarrow 0$, the 
coefficients $\Gamma_s^i$ tend to $2 \gamma$. The symmetric states within 
the subspace $\mc{S}$ display thus superradiant features since they decay faster 
as compared to two independent atoms.

\subsection{Non-degenerate System \label{non_degenerate}}
Here we discuss the diagonalization of   $H_A+H_{\Omega}$ 
in the most general case where the Zeeman splitting $\delta$ 
of the excited states is different from zero. 
The matrix representation of this Hamiltonian 
with respect to the states  $\{\ket{\psi_a^1},\ket{\psi_a^2},\ket{\psi_a^3}\}$  
defined in Eq.~(\ref{psi_a}) reads 
\begin{align}
[H_A+H_{\Omega}]_{\mc{A}} & = \hbar
\left(
\begin{array}{l@{\hspace{0.3cm}}l@{\hspace{0.3cm}}l}
\omega_0 +\far &  \delta\cos\theta & 0  \\
\delta\cos\theta& \omega_0+\far  & -\delta\sin\theta \\
0 & -\delta\sin\theta & \omega_0 + \near
\end{array}
\right)\,.
\label{H0HomegaA}
\end{align}
\begin{figure*}[t!] 
\includegraphics[scale=1]{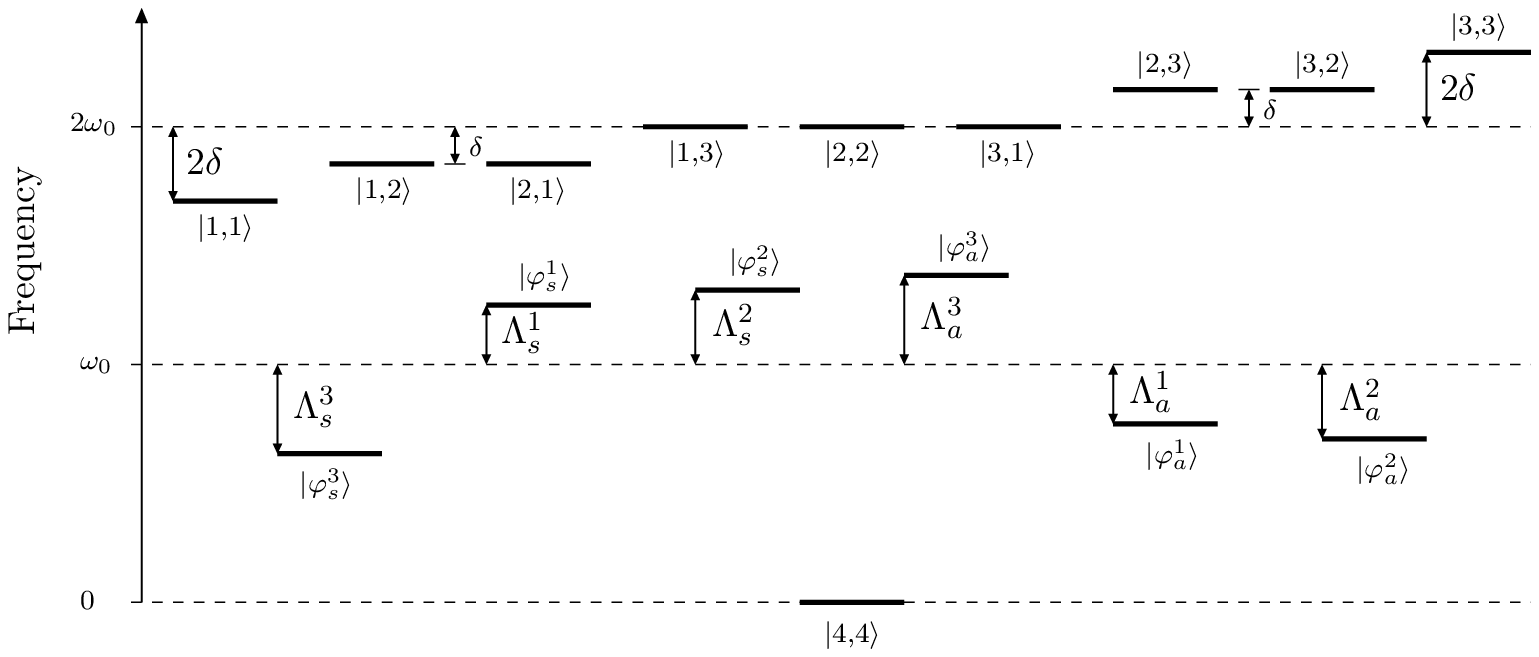}
\caption{\label{picture7} \small  
Complete level scheme  of the non-degenerate system ($\delta\not=0$).  
For the special geometrical setup 
where the atoms are aligned in the $x$-$y$-plane ($\theta=\pi/2$), the analytical expressions for the states 
$\ket{\vfi_a^i}$, $\ket{\vfi_s^i}$ and the frequency shifts   $\Lambda_a^i$, $\Lambda_s^i$ 
are given in Eqs.~(\ref{states_a}), (\ref{states_s}), (\ref{lambda_a}) and (\ref{lambda_s}), respectively. 
The frequency shifts  $\Lambda_a^i$  ($\Lambda_s^i$) of the antisymmetric (symmetric) states and  
the splitting of the excited states   are not to scale. 
Note that   the frequency shifts $\Lambda_a^i$  and $\Lambda_s^i$ 
depend on the relative position of the  atoms.  
} 
\end{figure*}
In general, the eigenvalues of this 
matrix can be written in the form 
\ba
E_a^1 & = & \hbar\left( \omega_0 + \Lambda_a^1 \right) \,, \nonumber \\
E_a^2 & = & \hbar \left( \omega_0 + \Lambda_a^2 \right) \,, \nonumber \\
E_a^3 & = & \hbar \left( \omega_0 + \Lambda_a^3 \right) \,,
\label{energy_a}
\ea
where the frequency shifts $\Lambda_a^i$ depend only on the interatomic 
distance $R$ and the azimuthal angle $\theta$, but not on the angle $\phi$. 
To illustrate this result, we consider a plane spanned by $\mf{e}_z$ 
and $\mf{e}_{\phi}=(\cos\phi,\sin\phi,0)$, see Fig.~\ref{picture5}. 
Within this plane, the vector $\mf{R}= z\,\mf{e}_z + l\, \mf{e}_{\phi}$ is 
described by the parameters $z$ and $l$, and 
Fig.~\ref{picture6}(a)-(c) shows  $\Lambda_a^i(l,z)$ as a function 
of these variables. Since the $\Lambda_a^i$ do not depend on $\phi$, the 
energy surfaces shown in Fig.~\ref{picture6}(a)-(c) remain the same  
if $\mf{e}_{\phi}$ is rotated around the $z$-axis. 
This result follows from the fact that the Hamiltonian $H_A$ in Eq.~(\ref{H0}) is invariant 
under rotations around the $z$- axis~\cite{kiffner:06}.

In Sec.~\ref{evolution}, we will focus on the geometrical setup where 
the atoms are aligned in the $x$- $y$- plane ($\theta=\pi/2$). 
In this case,  the frequency shifts $\Lambda_a^i$ of the antisymmetric states are 
found to be 
\ba
\Lambda_a^1 & = & \far           \,,                                      \nonumber \\[0.1cm]
\Lambda_a^2 & = & (\far +\near)/2 -  \omega_B/2 \,,\nonumber \\[0.1cm]
\Lambda_a^3 & = & (\far +\near)/2 + \omega_B/2 \,,  
\label{lambda_a}
\ea
where the Bohr frequency is given by 
\be
\omega_B   =   \sqrt{4 \delta^2+(\far -\near)^2} \,.
\label{rabi}
\ee
A plot of the frequency shifts $\Lambda_a^i$ as a function of the 
interatomic distance $R$ and for $\theta=\pi/2$ is shown in Fig.~\ref{picture6}(d).   
Note that the degeneracy and the level crossing of the eigenvalues $\lambda_a^i$ is 
removed for $\delta\not=0$ [see  Sec.~\ref{degenerate}]. 
The  eigenstates  that correspond to the frequency shifts  in Eq.~(\ref{lambda_a}) read 
\ba 
\ket{\vfi_a^1} & = &   \ket{a_2} \,, \nonumber \\[0.2cm]
\ket{\vfi_a^2} & = &  e^{i\xi}\sin\vartheta_a \ket{\psi_a^+} + \cos\vartheta_a \ket{\psi_a^-} \,,\nonumber \\[0.2cm]
 \ket{\vfi_a^3}  & = & - e^{i\xi}\cos\vartheta_a \ket{\psi_a^+} + \sin\vartheta_a \ket{\psi_a^-}   \,,
 \label{states_a}
\ea
where $\delta=|\delta|e^{i\xi}$  ($\xi\in \{0,\pi\}$), the states $\ket{\psi_a^{\pm}}$ are defined  
in Eq.~(\ref{psi_a_pm}), and the angle $\vartheta_a$ is determined by 
\be
\tan 2\vartheta_a = \frac{|\delta|}{\far -\near}\,,\quad 0 < \vartheta_a < \frac{\pi}{2}\,.
\ee
If the distance between the atoms is small such that $R\lesssim 0.63\times\lambda_0$, 
we have $\far<\near$. 
In this case,  we find 
$\lim_{\delta\rightarrow 0}\ket{\vfi_a^i}=\ket{\psi_a^i}$ 
and $\lim_{\delta\rightarrow 0}\Lambda_a^i=\lambda_a^i$, where 
the eigenstates $\ket{\psi_a^i}$  and the  
frequency shifts  $\lambda_a^i$ of the 
degenerate system are defined in Eqs.~(\ref{psi_a}) and (\ref{eigen_a}), respectively.  
 
The matrix representation of  $H_A+H_{\Omega}$  
with respect to the symmetric states   $\{\ket{\psi_s^1},\ket{\psi_s^2},\ket{\psi_s^3}\}$  
defined in Eq.~(\ref{psi_s}) is found to be 
\begin{align}
[H_A+H_{\Omega}]_{\mc{S}} & = \hbar
\left(
\begin{array}{l@{\hspace{0.3cm}}l@{\hspace{0.3cm}}l}
\omega_0 -\far &  \delta\cos\theta & 0  \\
\delta\cos\theta& \omega_0-\far  & -\delta\sin\theta \\
0 & -\delta\sin\theta & \omega_0 - \near
\end{array}
\right)\,.
\end{align}
Just as in the case of the antisymmetric states, 
the eigenvalues of  $[H_A+H_{\Omega}]_{\mc{S}}$ are written as 
\ba
E_s^1 & = & \hbar\left( \omega_0 + \Lambda_s^1 \right) \,, \nonumber \\
E_s^2 & = & \hbar \left( \omega_0 + \Lambda_s^2 \right) \,, \nonumber \\
E_s^3 & = & \hbar \left( \omega_0 + \Lambda_s^3 \right) \,, 
\ea
and the frequency shifts $\Lambda_s^i$  depend only on  the interatomic 
distance $R$ and the azimuthal angle $\theta$. 

If the atoms are aligned in the $x$- $y$- plane ($\theta=\pi/2$), 
the frequency shifts $\Lambda_s^i$ of the symmetric states  are 
given by 
\ba
\Lambda_s^1 & = & - \far           \,,                                      \nonumber \\[0.1cm]
\Lambda_s^2 & = & -(\far +\near)/2 +  \omega_B/2 \,,\nonumber \\[0.1cm]
\Lambda_s^3 & = & -(\far +\near)/2 - \omega_B/2 \,,  
\label{lambda_s}
\ea
and the corresponding eigenstates are 
\ba 
\ket{\vfi_s^1} & = &   \ket{s_2} \,, \nonumber \\[0.2cm]
\ket{\vfi_s^2} & = & - e^{i\xi}\cos\vartheta_s \ket{\psi_s^+} + \sin\vartheta_s \ket{\psi_s^-} \,,    \nonumber \\[0.2cm]
\ket{\vfi_s^3} & = & e^{i\xi}\sin\vartheta_s \ket{\psi_s^+} + \cos\vartheta_s \ket{\psi_s^-}\,.
\label{states_s}
\ea
The states $\ket{\psi_s^{\pm}}$ are defined  in Eq.~(\ref{psi_s_pm}),
$\delta=|\delta|e^{i\xi}$  ($\xi\in \{0,\pi\}$),  and the angle $\vartheta_s$ is determined by 
\be
\tan 2\vartheta_s = \frac{|\delta|}{\near  -  \far }\,,\quad 0 < \vartheta_s < \frac{\pi}{2}\,.
\ee
For small values of the 
interatomic distance $R$ such that $\far<\near$, 
 we find $\lim_{\delta\rightarrow 0}\ket{\vfi_s^i}=\ket{\psi_s^i}$ 
and $\lim_{\delta\rightarrow 0}\Lambda_s^i=\lambda_s^i$, where 
the eigenstates $\ket{\psi_s^i}$  and the  
frequency shifts  $\lambda_s^i$ of the 
degenerate system are defined in Eqs.~(\ref{psi_s}) and (\ref{eigen_s}), respectively.  

Finally, we note that the ground state $\ket{4,4}$ and 
the excited states $\ket{i,j}$ ($i,j \in\{1,2,3\}$) are eigenstates of $H_A+H_{\Omega}$. 
These states together with the symmetric and antisymmetric eigenstates of $H_A+H_{\Omega}$ 
form the new basis of the total state space  $\mc{H}_{\text{sys}}$.  
The complete level scheme of the non-degenerate system is shown in Fig.~\ref{picture7}.

\section{POPULATION OF THE DECOHERENCE FREE SUBSPACE \label{population}}
In this section we describe  a method that allows to 
populate the subspace $\mc{A}$ spanned by the antisymmetric states. 
For simplicity, we restrict the analysis to the degenerate system ($\delta=0$) 
and show how the states $\ket{\psi_a^i}$ 
can be populated selectively 
by means of an external laser field. 
However, a laser  field cannot induce direct transitions 
between the ground state $\ket{4,4}$ and  
$\ket{\psi_a^i}$ as long as  the electric 
field at the position of atom 1 is identical to the field at the location 
of atom 2. By contrast, a direct driving of the antisymmetric 
states is possible provided that one can realize a field gradient 
between the positions of the two atoms. Since we consider an interatomic spacing 
$R$ that is smaller than $\lambda_0/2$ such that the states in $\mc{A}$ are subradiant, 
the realization of this field gradient is an experimentally challenging task. 
Several authors proposed a setup where the atoms are 
placed symmetrically around the node of a standing light field~\cite{ficek:02,beige:00}, 
and this method also allows to address the states of our system individually. 
Other methods~\cite{akram:00,ficek:int,ficek:02} rest on the 
assumption that the atoms are \textit{non-identical} and cannot be applied 
to our system comprised of two identical atoms. 

\begin{figure}[t!]
\bc
\includegraphics[scale=1]{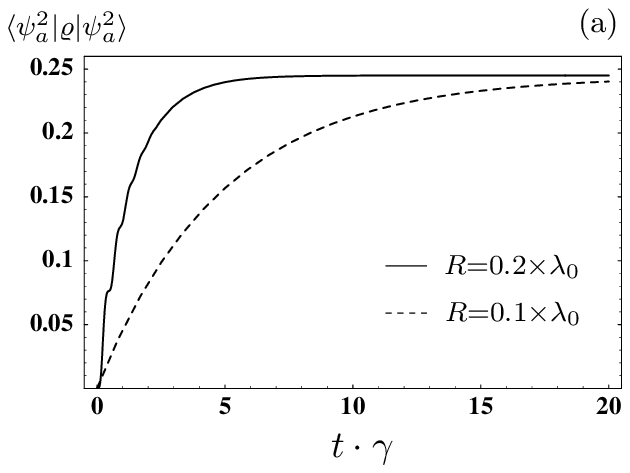}\\[0.3cm]
\includegraphics[scale=1]{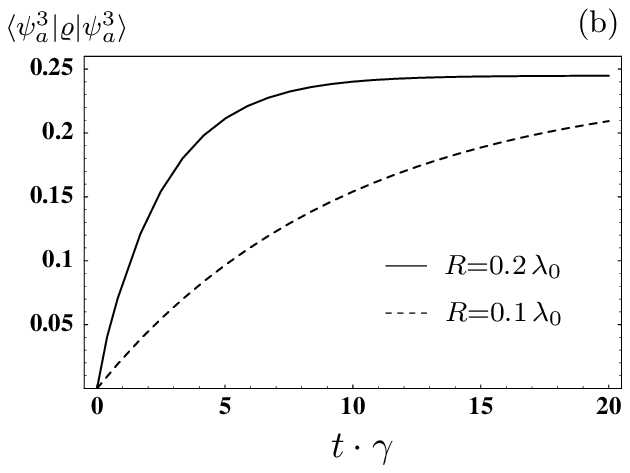}
\caption{\label{picture9} \small 
Time-dependent population of the states $\ket{\psi_a^i}$ for different polarizations  
of the driving field. The initial state at $t=0$ is $\ket{4,4}$. 
The parameters are $\theta=\pi/2$, $\phi=0$, $\delta=0$ and $\Delta_2=0$. 
(a) Population of $\ket{\psi_a^2}$ for  $\Omega_y(\mf{r}_1)=\Omega_y(\mf{r}_2)= 5\gamma$. 
The states $\ket{\psi_a^1}$ and $\ket{\psi_a^3}$ are not populated. 
(b) Population of $\ket{\psi_a^3}$ for   $\Omega_x(\mf{r}_1)=\Omega_x(\mf{r}_2)= 5\gamma$. 
The states $\ket{\psi_a^1}$ and $\ket{\psi_a^2}$ are not populated.
  }
\ec
\end{figure}

\begin{table}[b!]
\begin{tabular}{l@{\hspace{0.2cm}}||@{\hspace{0.2cm}}c@{\hspace{0.2cm}}|@{\hspace{0.2cm}}
c@{\hspace{0.2cm}}|@{\hspace{0.2cm}}c@{\hspace{0.2cm}}|}
& $\ket{\psi_a^1}  $                                               &   $\ket{\psi_a^2}$          &  $\ket{\psi_a^3}$ \\[0.2cm]
\hline\hline
$\ket{1,2}$ & $\mf{e}_x,\,  \mf{e}_y $  &  $ \mf{e}_z$           &   $\mf{e}_z $ \\[0.2cm]
\hline
$\ket{2,1}$ & $\mf{e}_x,\,  \mf{e}_y$   &   $\mf{e}_z $          &  $ \mf{e}_z$     \\[0.2cm]
\hline
$\ket{1,3}$ & -                                                         &   $\mf{e}_x  $       &    $ \mf{e}_y$           \\[0.2cm]
\hline
$\ket{3,1}$ & -                                                         &   $\mf{e}_x $          &   $\mf{e}_y$  \\[0.2cm]
\hline
$\ket{2,3}$ & $\mf{e}_x,\,  \mf{e}_y$   &    $\mf{e}_z $         &    $\mf{e}_z$ \\[0.2cm]
\hline
$\ket{3,2}$ & $\mf{e}_x,\,  \mf{e}_y $   &   $\mf{e}_z  $         &    $\mf{e}_z $          \\[0.2cm]
\hline
\end{tabular}
\caption{\small \label{coupling}
Polarization of the external laser field that couples an antisymmetric state $\ket{\psi_a^i}$ to 
an excited  state $\ket{i,j}$ ($i,j\in\{1,2,3\}$)  for $\delta=0$.  
Note that $\ket{\psi_a^1}$ does not couple to $z$-polarized light, $\ket{\psi_a^2}$ 
does not couple to $y$-polarized light and  $\ket{\psi_a^1}$ does not couple to 
$x$-polarized light. See also Fig.~\ref{picture8}. 
}
\end{table}
\begin{figure}[t!]
\bc
\includegraphics[scale=1]{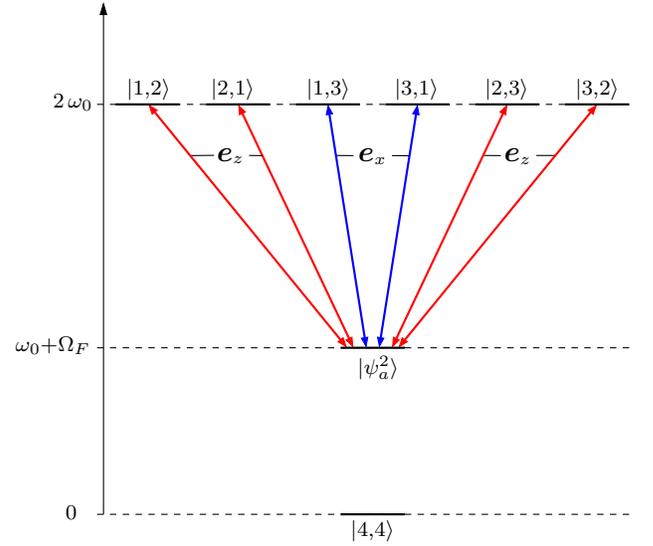}
\caption{\label{picture8} \small (Color online) 
Laser-induced coupling of $\ket{\psi_a^2}$ to the excited states $\ket{i,j}$ ($i,j\in\{1,2,3\}$) 
in the case of the degenerate system.  
States that are not directly coupled to $\ket{\psi_a^2}$  
have been omitted (except for the ground state).  
The laser  polarization that  couples the antisymmetric state $\ket{\psi_a^2}$ to 
a state $\ket{i,j}$ ($i,j\in\{1,2,3\}$)  is indicated next to the respective transition.  
$\ket{\psi_a^2}$ is completely decoupled from a $y$-polarized laser field.  
  }
\ec
\end{figure}
Here we describe a  method that allows to populate the states 
$\ket{\psi_a^i}$ individually and that does not require a field gradient 
between the positions of the two atoms. It rests on a finite distance between 
the atoms and exploits the fact that the antisymmetric states 
may be  populated by spontaneous emission from the 
excited states $\ket{i,j}$ ($i,j\in\{1,2,3\}$). 
For a given  geometrical setup, we 
choose a coordinate system where the  unit vector $\mf{e}_x$ 
coincides with the separation vector $\mf{R}$. 
In this case, we have $\theta=\pi/2$ and $\phi=0$. 
The  $z$-direction  is determined by the external magnetic field 
and can be chosen in any direction perpendicular to $\mf{R}$. 
The polarization vector of the laser field propagating in 
$z$-direction lies in the $x$-$y$-plane 
and can be adjusted as needed, see Eq.~(\ref{laser_field}).
In the presence of the laser, the atomic evolution is governed by the 
master equation~(\ref{master_L}). 
We find that the coupling of the states $\ket{\psi_a^i}$ to 
the excited states $\ket{i,j}$ ($i,j\in\{1,2,3\}$) depends on 
the polarization of the laser field 
(see Table~\ref{coupling} and  Fig.~\ref{picture8}). 
In particular, it is found  that $\ket{\psi_a^1}$ does not couple to $z$-polarized light, 
$\ket{\psi_a^2}$ does not couple to $y$-polarized light and $\ket{\psi_a^3}$ does not 
couple to $x$-polarized light. At the same time, the states $\ket{\psi_a^i}$ are populated 
by spontaneous emission from the excited states. 
This fact together with the polarization dependent coupling 
of the antisymmetric states allows to populate the states $\ket{\psi_a^i}$ selectively. 
In order to populate state $\ket{\psi_a^2}$, for example, one has to shine in a $y$-polarized 
field. Since the spontaneous decay of  $\ket{\psi_a^2}$ is slow and since $\ket{\psi_a^2}$ is decoupled 
from the laser, population can accumulate in this state. On the other 
hand, the states $\ket{\psi_a^1}$ and $\ket{\psi_a^3}$ are depopulated by the 
laser coupling to the excited states. This situation is shown in 
Fig.~\ref{picture9}(a) for two different values of the interatomic distance $R$. 
The initial state at $t=0$ is $\ket{4,4}$, and for $  t\cdot\gamma=20$ the population 
of $\ket{\psi_a^2}$ is approximately 1/4. Since all coherences between $\ket{\psi_a^2}$ and 
any other state are zero, the probability to find the system at $t= 20 /\gamma $ in the 
pure state $\ket{\psi_a^2}$ is given by 1/4.

The exact steady state solution of Eq.~(\ref{master_L}) is difficult to obtain analytically. 
However, one can determine the steady state value of  $\bra{\psi_a^2}\vro\ket{\psi_a^2}$ 
with the help of Eq.~(\ref{eom_anti}), 
\be
\bra{\psi_a^2}\vro_{\text{st}}\ket{\psi_a^2} =\left[\lim\limits_{t\rightarrow\infty}C_a^2(t)\right]/( 2 \Gamma_a^2)\,.
\ee
The population of $\ket{\psi_a^2}$ in steady state is thus limited by the population of the relevant excited states 
that are populated by the $y$-polarized laser field and that decay spontaneously to $\ket{\psi_a^2}$. 
Furthermore, it is possible to gain some insight into the time evolution of $\bra{\psi_a^2}\vro\ket{\psi_a^2}$. 
For a strong laser field and for a small value of $R$, $C_a^2$ reaches the steady state on a timescale that is 
fast as compared to $1/(2\Gamma_a^2)$. We may thus replace $C_a^2$ by its steady state value in 
Eq.~(\ref{eom_anti}). The solution of this differential equation is 
\be
\bra{\psi_a^i}\vro(t)\ket{\psi_a^i} \approx \frac{\left[\lim\limits_{t\rightarrow\infty}C_a^i(t)\right]}{ 2 \Gamma_a^i}
\left[1-e^{-2\Gamma_a^2 t}\right]
\label{time_pop}
\ee
and reproduces the exact time evolution of $\bra{\psi_a^2}\vro\ket{\psi_a^2}$ according to Fig.~\ref{picture9}(a) 
quite well. 
Moreover, it becomes now clear why it takes longer until the 
population of  $\ket{\psi_a^2}$ reaches its steady state if the interatomic distance $R$ is reduced since  
the decay rate $2\Gamma_a^2$ approaches zero as $R\rightarrow 0$.

So far, we considered only the population of $\ket{\psi_a^2}$, 
but the treatment of $\ket{\psi_a^1}$ and $\ket{\psi_a^3}$ is 
completely analogous. The population of $\ket{\psi_a^3}$ by a $x$-polarized field  
is shown  in Fig.~\ref{picture9}(b). The differences between plot (a) and (b) arise since 
the decay rates of $\ket{\psi_a^2}$ and $\ket{\psi_a^3}$ are different for the 
same value of $R$ (see Sec.~\ref{section_decay}). 
In general, the presented method may also  be employed 
to populate the antisymmetric states of the non-degenerate system selectively. 
In this case, the polarization of the field needed to populate a state $\ket{\vfi_a^i}$ is 
a function of the detuning $\delta$.

In conclusion, the discussed  method  allows to populate  the antisymmetric states selectively, 
provided that  the interatomic distance is larger than zero. 
If the interatomic distance is reduced, a longer interaction time with the laser field 
is required to reach the maximal value of $\bra{\psi_a^i}\vro\ket{\psi_a^i}\approx1/4$. 
Note that a finite distance between the atoms is also required 
in the case of other schemes where the atoms are 
placed symmetrically around the node of a standing light field~\cite{ficek:02,beige:00}.  
While the latter method allows, at least in  principle, for a complete population transfer to 
the antisymmetric states, its experimental  realization is difficult for two nearby atoms. 
By contrast,   our scheme  does not require a field gradient between the atoms and is 
thus easier to implement. 
It has been pointed out that the  population transfer to the antisymmetric states  
is limited by the population of the excited states 
that spontaneously decay to an antisymmetric state $\ket{\psi_a^i}$. 
Although this limit is difficult to overcome, an improvement can be  achieved if the 
fluorescence intensity is observed while the atom is 
irradiated by the laser.  
As soon as the system decays into one of the states $\ket{\psi_a^i}$,  
the fluorescence signal is interrupted for a time period that is on the order of $1/(2\Gamma_a^i)$ 
(see Sec.~\ref{section_decay}). The dark periods in the fluorescence signal reveal thus the 
spontaneous emission events that lead to the population of one of  the antisymmetric states.

\section{INDUCING DYNAMICS WITHIN THE SUBSPACE  $\mc{A}$ \label{evolution}}
\begin{figure}[t!]
\bc
\includegraphics[scale=1]{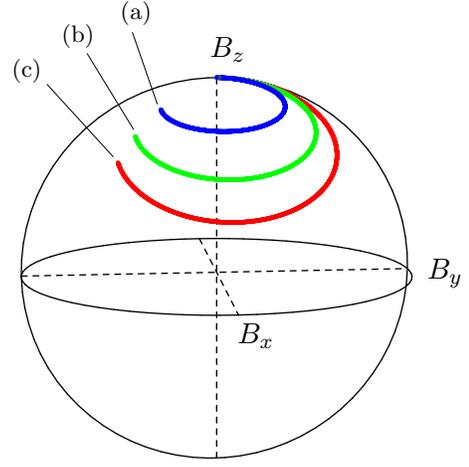} 
\caption{\label{picture10} \small (Color online) 
Bloch sphere representation of the system dynamics in the subspace $\mc{Q}$ 
spanned by the states $\{\ket{\psi_a^2},\,\ket{\psi_a^3}\}$. At $t=0$, the system is 
in the pure state $\ket{\psi_a^2}$ and a static magnetic 
field is switched on. The Bloch vector is rotated 
around an axis in the $x$-$z$-plane, and the tilt of this axis in $x$-direction increases 
with the magnetic field strength. The value of the parameter $\delta$ 
is (a) $\delta =3.15\times \gamma $, (b) $\delta =4.83\times\gamma $ and 
(c) $\delta = 6.22\times\gamma$, and we chose $R= 0.1\times \lambda_0$. 
  }
\ec
\end{figure}

In this Section we assume that the system has been prepared in the antisymmetric state 
$\ket{\psi_a^2}$, for example by one of the methods described in Sec.~\ref{population}. 
The aim is to induce a  controlled dynamics in the subspace $\mc{A}$ of the antisymmetric states. 
We suppose that the atoms are aligned along the $x$-axis, i.e. $\theta=\pi/2$ and $\phi=0$. 
According to Eq.~(\ref{H0HomegaA}), the state $\ket{\psi_a^2}$ is then only coupled to $\ket{\psi_a^3}$. 
Apart from a  constant, the Hamiltonian $H_{\mc{Q}} $ 
that governs the unitary time evolution in the space $\mc{Q}$ spanned by 
$\{\ket{\psi_a^2},\,\ket{\psi_a^3}\}$  can be written as 
\begin{align}
H_{\mc{Q}}  &  = \hbar
\left(
\begin{array}{l@{\hspace{0.3cm}}l }
-(\near  -\far)/2 &  -  \delta  \\
 - \delta  & ( \near - \far)/2
\end{array}
\right) \notag \\[0.4cm]
 & = \hbar \omega_B \mf{\hat{n}}\cdot \mf{\sigma} /2\,,
 \label{H_static}
\end{align} 
where the vector 
$\mf{\sigma}=\{\sigma_x,\sigma_y,\sigma_z\}$ consists of  the Pauli matrices 
$\sigma_i$, and the unit vector $\mf{\hat{n}}$ is defined as 
\be
\mf{\hat{n}} = -\left( 2\delta  ,0,  \near-\far  \right)/ \omega_B\,.
\label{axis}
\ee
The Bohr frequency $\omega_B$ is the difference between the eigenvalues of $H_{\mc{Q}}$ and is 
given in Eq.~(\ref{rabi}) of Sec.~\ref{non_degenerate}.  
Equation~(\ref{H_static}) implies that  the parameter $\delta$ which can be adjusted 
by means of the external magnetic field introduces a coupling between the states 
$\ket{\psi_a^2}$ and $\ket{\psi_a^3}$.   
If the initial state is $\ket{\psi_a^2}$, the final state $\ket{\psi_F}$ reads
\be
\ket{\psi_F(t)} = U(t,0)\ket{\psi_a^2},\,
\ee
where $U=\exp(-i H_{\mc{Q}}t/\hbar)$ is the time evolution operator. 
The  time evolution induced by $H_{\mc{Q}}$  can be described in a 
simple way in the Bloch sphere picture~\cite{nielsen:00}. 
The Bloch vector of the state $\ket{\psi_F(t)}$  is  defined as 
\be
\mf{B}(t) = \bra{\psi_F(t)}\mf{\sigma}\ket{\psi_F(t)}\,.
\ee
Initially, this vector points into the positive $z$-direction. The time evolution 
operator $U$ rotates this vector on the Bloch sphere around the axis $\mf{\hat{n}}$ 
by an angle $\omega_B t$. 
According to Eq.~(\ref{axis}), the axis of rotation lies in the $x$-$z$-plane and 
its orientation depends on the parameter $\delta$ which can be controlled 
by means of the magnetic field. 
In order to demonstrate these analytical considerations, we numerically integrate the 
master equation~(\ref{master}) with the initial condition $\vro(t=0)=\ket{\psi_a^2}\bra{\psi_a^2}$. 
We define a projector onto the space spanned by $\{\ket{\psi_a^2},\,\ket{\psi_a^3}\}$, 
\be
\hat{P}=\ket{\psi_a^2}\bra{\psi_a^2} + \ket{\psi_a^3}\bra{\psi_a^3}\,.
\ee
The generalized Bloch vector is then defined as 
\be
\mf{B}_N(t) = \text{Tr}\left[\mf{\sigma}\hat{P}\vro(t)\hat{P}\right]\,.
\label{bloch_gen}
\ee
In contrast to $\mf{B}$, $\mf{B}_N$ is not necessarily a unit vector, but its length can 
be smaller than unity due to spontaneous emission from $\ket{\psi_a^2}$  and $\ket{\psi_a^3}$ to the ground state.  
Figure~\ref{picture10} shows the evolution of $\mf{B}_N$ for different values of the 
parameter $\delta$ which depends on the magnetic field strength. 
Let  $\mf{S}=\{S_x,S_y,S_z\}$ be a point on the Bloch sphere 
that lies not in the $y$-$z$-plane ($S_x\not=0$). 
If one chooses the parameter $\delta$  according to 
\be
\delta   =  \frac{1-S_z}{2|S_x|}|\far -\near|\;\text{Sign}(S_x)  \,, 
\label{delta}
\ee
then  $\mf{S}$ lies on the orbit of  the rotating  Bloch vector  $\mf{B}$ if 
spontaneous emission is negligible. According to Eq.~(\ref{delta}), 
any point close to the $y$-$z$-plane requires large values of $\delta$ 
since $|\delta|$ diverges for $S_x\rightarrow 0$. 
The dynamics that can be induced by  a static magnetic 
field is thus restricted, particularly because 
we are only considering the regime of the linear Zeeman effect. 

These limitations can be overcome if a radio-frequency (RF) field 
is applied instead of a static magnetic field. 
If the RF field oscillates along the $z$-axis, the Hamiltonian 
$H_A$ in Eq.~(\ref{H0}) has to be replaced by 
\be
H_A^{\text{rf}}(t)   =   \hbar \omega_0\sum\limits_{i=1}^3  \sum\limits_{\mu=1}^{2} \, 
\Sp{i}{\mu} \Sm{i}{\mu} + V_{\text{rf}}(t) \,, 
\label{H_A_mic}
\ee
where
\be
V_{\text{rf}}(t) =2  \hbar \delta(t)
 \sum\limits_{\mu=1}^{2} \, \left(\Sp{3}{\mu} \Sm{3}{\mu}-\Sp{1}{\mu} \Sm{1}{\mu}\right)
\ee
 describes the interaction with the RF field and 
\be
\delta(t) =  \delta_0 \cos(\omega_{\text{rf}}\, t + \phi_{\text{rf}})\,.
\ee
In this equation, the magnitude of $\delta_0(>0)$ 
depends on the amplitude of the RF field, and  $\omega_{\text{rf}}$ and $\phi_{\text{rf}}$ are 
the  frequency and phase of the RF field, respectively. 
We assume that the interatomic distance of the atoms is smaller than $R= 0.63\times\lambda_0$.
In this case, the dipole-dipole interaction raises the 
energy of $\ket{\psi_a^3}$ with respect to $\ket{\psi_a^2}$, and the frequency 
difference between these two states is   $\near-\far>0$. 
Furthermore, we suppose that 
the detuning  $\Delta_{\text{rf}}=\omega_{\text{rf}}-(\near-\far )$ of the 
RF field with the  $\ket{\psi_a^2}\leftrightarrow\ket{\psi_a^3}$ transition and the 
parameter $\delta_0$  are  
small as compared to $(\near-\far )$ such that the 
rotating-wave approximation  can be employed. 
In a frame rotating with   $\omega_{\text{rf}}$,  
the system dynamics in the subspace $\mc{Q}$ spanned by $\{\ket{\psi_a^2},\,\ket{\psi_a^3}\}$ 
is then governed by the Hamiltonian
\begin{align}
H_{\mc{Q}}^{\text{rf}}  &  =  \hbar 
\left(
\begin{array}{l@{\hspace{0.3cm}}l }
\Delta_{\text{rf}}/2  &  -  \delta_0 \exp(i\phi_{\text{rf}})  \\[0.2cm]
 - \delta_0 \exp(-i\phi_{\text{rf}})  & -\Delta_{\text{rf}}/2
\end{array}
\right) \notag \\[0.4cm]
 & = \hbar \Omega_{\text{rf}}\, \mf{\hat{n}}_{\text{rf}}\cdot \mf{\sigma} /2\,, 
 \label{H_micro}
\end{align} 
where 
\be
\mf{\hat{n}}_{\text{rf}} = \left( - 2\delta_0 \cos\phi_{\text{rf}}  ,
- 2\delta_0 \sin \phi_{\text{rf}} ,  \Delta_{\text{rf}}   \right)/\Omega_{\text{rf}}
\label{axis2}
\ee
and $\Omega_{\text{rf}}=\sqrt{\Delta_{\text{rf}}^2+ 4 |\delta_0|^2}$.  
For a resonant RF field ($\Delta_{\text{rf}}=0$), the axis $\mf{\hat{n}}_{\text{rf}}$ 
lies in the $x$-$y$-plane of the Bloch sphere, and its orientation can be adjusted 
at will by the phase $\phi_{\text{rf}}$ of the RF field. 
Any single-qubit operation can thus be realized by a sequence of suitable RF pulses~\cite{nielsen:00}. 
In particular, a complete transfer of population from $\ket{\psi_a^2}$ to $\ket{\psi_a^3}$ 
can  be achieved by a resonant RF pulse with a duration of $\pi /\Omega_{\text{rf}}$ 
and an arbitrary phase $\phi_{\text{rf}}$.  
\begin{figure}[t!]
\includegraphics[scale=1]{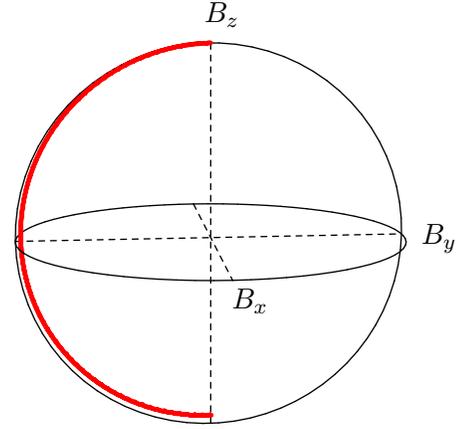}
\caption{\label{picture11} \small (Color online) 
Complete population transfer from  $\ket{\psi_a^2}$ to $\ket{\psi_a^3}$ 
by means of a resonant RF field. 
At $t=0$, the Bloch vector $\mf{B}_N$ points into the positive $z$ direction. At $t=\pi/\Omega_{\text{rf}}$, the 
state of the system is $\ket{\psi_a^3}$ and $\mf{B}_N$ points into the negative $z$ direction. 
Note that the length of  $\mf{B}_N$ is slightly smaller than unity for $t>0$ due to the 
small probability of spontaneous emission to the ground state. The parameters are 
$R= 0.05\times \lambda_0$, $\delta_0 = \gamma$, $ \phi_{\text{rf}} =\pi $ and $\Delta_{\text{rf}} = 0$. 
  }
\end{figure}

Next we demonstrate that the Hamiltonian $H_{\mc{Q}}^{\text{rf}}$ in 
Eq.~(\ref{H_micro}) describes the system dynamics quite 
well if the atoms are close to each other such that spontaneous emission 
is strongly suppressed. For this, 
we  transform the master equation~(\ref{master}) with 
$H_A^{\text{rf}}$  instead of   $H_A$   in 
a frame rotating with $\omega_{\text{rf}}$. The resulting 
equation is integrated numerically without making   
the  rotating-wave approximation. 
We suppose that the system is initially in the state $\ket{\psi_a^2}$, 
and the phase of the resonant RF field has been set to $\phi_{\text{rf}}= \pi$. 
Figure~\ref{picture11} shows the time evolution of the 
Bloch vector $\mf{B}_N$. As predicted by Eq.~(\ref{H_micro}), the 
Bloch vector is rotated around the $x$-axis and 
at $t=\pi/\Omega_{\text{rf}}$, 
$\mf{B}_N$ points in the negative $z$-direction. Due to the 
small probability of spontaneous emission to the 
ground state, the length of $\mf{B}_N$ 
is slightly smaller than unity ($|\mf{B}_N|=0.95$) at $t=\pi/\Omega_{\text{rf}}$.   

Finally, we briefly discuss how the final state $\ket{\psi_F(t)}$ could be 
measured. In principle, one can exploit the polarization-dependent coupling 
of the states  $\ket{\psi_a^2}$ and $\ket{\psi_a^3}$ to the excited states 
(see Sec.~\ref{population}). For example, one could ionize the 
system in a two-step process, where    $\ket{\psi_a^2}$ 
($\ket{\psi_a^3}$) is first resonantly coupled to the excited states $\ket{i,j}$  
($i,j\in\{1,2,3\}$). A second laser then ionizes the system, 
and the ionization rate is a measure for the population of state 
$\ket{\psi_a^2}$  ($\ket{\psi_a^3}$). Another possibility 
is to shine in a single laser whose 
frequency is just high enough to ionize the system 
starting from $\ket{\psi_a^3}$. Since 
the energy of $\ket{\psi_a^3}$  is higher than those of 
$\ket{\psi_a^2}$, the ionization rate is a measure for 
the population of state $\ket{\psi_a^3}$. 

\section{ENTANGLEMENT OF THE COLLECTIVE TWO-ATOM STATES \label{entanglement}}
In Sec.~\ref{degenerate}, we determined the collective two-atom states 
$\ket{\psi_a^i}$ and $\ket{\psi_s^i}$ that are formed by the 
coherent part of the dipole-dipole interaction. Here we show that
these states are entangled, i.e. they cannot be written as a single tensor product 
$\ket{\psi_1}\otimes\ket{\psi_2}$ of two single-atom states. 
In order to quantify the degree of entanglement, 
we calculate the concurrence~\cite{wooters:98,rungta:01}  of the 
pure states $\ket{\psi_a^i}$ and $\ket{\psi_s^i}$. 
The concurrence for a pure state $\ket{\psi_{12}}$ 
of the two-atom state space $\mc{H}_{\text{sys}}=
\mc{H}_1\otimes \mc{H}_2$ is defined as~\cite{rungta:01}
\be
C(\ket{\psi_{12}}) = \sqrt{2[1 - \text{Tr}  (\vro_1^2)]} \,.
\ee
Here $\vro_1 = \text{Tr}_2(\vro)$ denotes the reduced density operator 
of atom 1. The concurrence $C$ of a maximally entangled state 
in $\mc{H}_{\text{sys}}$ is $C_{\text{max}}=\sqrt{3/2}$, 
and $C$ is zero for product states~\cite{rungta:01}. 
We find that the antisymmetric and symmetric states $\ket{\psi_a^i}$ and 
$\ket{\psi_s^i}$ are   entangled,  but the degree of entanglement 
is not maximal,
\be
C(\ket{\psi_a^i}) = C(\ket{\psi_s^i}) = 1 < C_{\text{max}} \, .
\ee
Next we compare this result to the corresponding results for 
a pair of interacting two-level systems with 
ground state $\ket{g}$  and excited state $\ket{e}$. 
In this case,   the exchange interaction gives rise to the   
entangled states~\cite{dicke:54,ficek:02,ficek:int}
\be
\ket{ \pm } = \frac{1}{\sqrt{2}}(\ket{e,g}\pm \ket{g,e})
\ee
with $C(\ket{ \pm }) = 1$. 
It follows that the degree of entanglement of the  states 
$\ket{\pm}$ is the same than the 
degree of entanglement of the  symmetric and antisymmetric states  
of two  four-level systems. 
On the other hand, the states $\ket{\pm}$ are maximally entangled in 
the state space of two two-level systems. This is in contrast to 
the states $\ket{\psi_a^i}$ and $\ket{\psi_s^i}$ which are 
not maximally entangled in the state space of two four-level atoms. 
Note that  the system of two four-level atoms shown in Fig.~\ref{picture1} 
may be reduced to a pair of two-level systems if 
the atoms are aligned along the $z$-axis. 
For this particular setup, all cross-coupling terms $\Omega_{ij}$ and 
$\Gamma_{ij}$ with $i\not=j$ vanish [see Eqs.~(\ref{OmegaExplicit}) and~(\ref{GammaExplicit})] 
such that an arbitrary sublevel of the $P_1$ triplet and the ground state $S_0$ 
form an effective two-level system. 

In Sec.~\ref{evolution}, we showed that a static magnetic or RF field can 
induce a controlled dynamics between the states $\ket{\psi_a^2}$ 
and $\ket{\psi_a^3}$. We find that the degree of entanglement 
of an arbitrary superposition state 
\be
\ket{\psi_{\text{sup}}} = a \ket{\psi_a^2} + b\ket{\psi_a^3}
\ee 
with $|a|^2 + |b|^2 =1$ is given by $C(\ket{\psi_{\text{sup}}}) = 1$. 
It follows that the degree of entanglement is not influenced 
by the induced dynamics between the states 
$\ket{\psi_a^2}$ and $\ket{\psi_a^3}$.

Finally, we point out that the antisymmetric states $\ket{\psi_a^i}$ 
can be populated selectively, for example by the method 
introduced in Sec.~\ref{population}. Since the spontaneous 
decay of the antisymmetric states is suppressed if the interatomic 
distance is small as compared to mean transition wavelength $\lambda_0$, 
we have shown that the system can be prepared in   long-lived 
entangled states. 

\section{SUMMARY AND DISCUSSION \label{discussion}}
We have shown that the state space of two dipole-dipole interacting 
four-level atoms contains a four-dimensional 
decoherence-free subspace (DFS) if 
the interatomic distance approaches zero. If the separation 
of the atoms is larger than zero but small as compared to the wavelength 
of the $S_0\leftrightarrow P_1$ transition, the spontaneous 
decay of states within the DFS is suppressed. 
In addition, we have shown that the system dynamics 
within the DFS is closed, i.e., the coherent part of the dipole-dipole interaction 
does not introduce a coupling between states of the DFS and 
states outside of the DFS. 

In the case of degenerate excited states ($\delta=0$), 
we find  that the  energy levels 
depend only on the interatomic distance $R$, but not on the 
angles $\theta$ and $\phi$.  
This result reflects the fact that each atom is modelled by complete 
sets of angular momentum multiplets~\cite{kiffner:06}. 
We identified two antisymmetric collective states 
($\ket{\psi_a^2}$ and $\ket{\psi_a^3}$) within the DFS that can be employed 
to represent a qubit. 
The storing times of the qubit state depend on the interatomic distance $R$ 
and can be significantly longer  than the inverse decay rate of 
the $S_0\leftrightarrow P_1$ transition.  
Moreover, any single-qubit operation can be 
realized via a sequence of suitable RF pulses. The energy splitting between the 
states $\ket{\psi_a^2}$ and  $\ket{\psi_a^3}$ arises from 
the coherent dipole-dipole interaction between the atoms and 
is on the order of $10 \gamma\equiv(10-1000)$ MHz
in the relevant interatomic distance range. 
The coupling strength between the RF field and the atoms  
is characterized by the parameter $\delta_0$ which is 
on the order of $\mu_B B_0$, where $\mu_B$ is 
the Bohr magneton and $B_0$ is the amplitude of the RF field.   
Since $\mu_B$ is about $3$ orders of magnitude larger than the 
nuclear magneton, typical operation times of our system may be  
significantly shorter than for a nuclear spin system.


\begin{thebibliography}{44}
\expandafter\ifx\csname natexlab\endcsname\relax\def\natexlab#1{#1}\fi
\expandafter\ifx\csname bibnamefont\endcsname\relax
  \def\bibnamefont#1{#1}\fi
\expandafter\ifx\csname bibfnamefont\endcsname\relax
  \def\bibfnamefont#1{#1}\fi
\expandafter\ifx\csname citenamefont\endcsname\relax
  \def\citenamefont#1{#1}\fi
\expandafter\ifx\csname url\endcsname\relax
  \def\url#1{\texttt{#1}}\fi
\expandafter\ifx\csname urlprefix\endcsname\relax\def\urlprefix{URL }\fi
\providecommand{\bibinfo}[2]{#2}
\providecommand{\eprint}[2][]{\url{#2}}

\bibitem[{\citenamefont{Chuang et~al.}(1995)\citenamefont{Chuang, Laflamme,
  Shor, and Zurek}}]{chuang:95}
\bibinfo{author}{\bibfnamefont{I.~L.} \bibnamefont{Chuang}},
  \bibinfo{author}{\bibfnamefont{R.}~\bibnamefont{Laflamme}},
  \bibinfo{author}{\bibfnamefont{P.~W.} \bibnamefont{Shor}}, \bibnamefont{and}
  \bibinfo{author}{\bibfnamefont{W.~H.} \bibnamefont{Zurek}},
  \bibinfo{journal}{Science} \textbf{\bibinfo{volume}{270}},
  \bibinfo{pages}{1633} (\bibinfo{year}{1995}).

\bibitem[{\citenamefont{Ekert and Jozsa}(1996)}]{ekert:96}
\bibinfo{author}{\bibfnamefont{A.}~\bibnamefont{Ekert}} \bibnamefont{and}
  \bibinfo{author}{\bibfnamefont{R.}~\bibnamefont{Jozsa}},
  \bibinfo{journal}{Rev. Mod. Phys.} \textbf{\bibinfo{volume}{68}},
  \bibinfo{pages}{733} (\bibinfo{year}{1996}).

\bibitem[{\citenamefont{Nielsen and Chuang}(2000)}]{nielsen:00}
\bibinfo{author}{\bibfnamefont{M.~A.} \bibnamefont{Nielsen}} \bibnamefont{and}
  \bibinfo{author}{\bibfnamefont{I.~L.} \bibnamefont{Chuang}},
  \emph{\bibinfo{title}{Quantum Computation and Quantum Information}}
  (\bibinfo{publisher}{Cambridge University Press},
  \bibinfo{address}{Cambridge}, \bibinfo{year}{2000}).

\bibitem[{\citenamefont{Monroe}(2002)}]{monroe:02}
\bibinfo{author}{\bibfnamefont{C.}~\bibnamefont{Monroe}},
  \bibinfo{journal}{Nature} \textbf{\bibinfo{volume}{416}},
  \bibinfo{pages}{238} (\bibinfo{year}{2002}).

\bibitem[{\citenamefont{DiVincenzo}(1995)}]{divincenzo:95}
\bibinfo{author}{\bibfnamefont{D.~P.} \bibnamefont{DiVincenzo}},
  \bibinfo{journal}{Science} \textbf{\bibinfo{volume}{270}},
  \bibinfo{pages}{255} (\bibinfo{year}{1995}).

\bibitem[{\citenamefont{Unruh}(1995)}]{unruh:95}
\bibinfo{author}{\bibfnamefont{W.~G.} \bibnamefont{Unruh}},
  \bibinfo{journal}{Phys. Rev. A} \textbf{\bibinfo{volume}{51}},
  \bibinfo{pages}{992} (\bibinfo{year}{1995}).

\bibitem[{\citenamefont{Zanardi and Rasetti}(1997)}]{zanardi:97}
\bibinfo{author}{\bibfnamefont{P.}~\bibnamefont{Zanardi}} \bibnamefont{and}
  \bibinfo{author}{\bibfnamefont{M.}~\bibnamefont{Rasetti}},
  \bibinfo{journal}{Phys. Rev. Lett.} \textbf{\bibinfo{volume}{79}},
  \bibinfo{pages}{3306} (\bibinfo{year}{1997}).

\bibitem[{\citenamefont{Lidar et~al.}(1998)\citenamefont{Lidar, Chuang, and
  Whaley}}]{lidar:98}
\bibinfo{author}{\bibfnamefont{D.~A.} \bibnamefont{Lidar}},
  \bibinfo{author}{\bibfnamefont{I.~L.} \bibnamefont{Chuang}},
  \bibnamefont{and} \bibinfo{author}{\bibfnamefont{K.~B.}
  \bibnamefont{Whaley}}, \bibinfo{journal}{Phys. Rev. Lett.}
  \textbf{\bibinfo{volume}{81}}, \bibinfo{pages}{2594} (\bibinfo{year}{1998}).

\bibitem[{lid()}]{lidar:dfs}
\bibinfo{note}{D. A. Lidar and K. B. Whaley, in \textit{Irreversible Quantum
  Dynamics}, edited by F. Benatti and R. Floreanini, (Springer Lecture Notes in
  Physics vol. \textbf{622}, Berlin, 2003), pp. 83-120.}

\bibitem[{\citenamefont{Kempe et~al.}(2001)\citenamefont{Kempe, Bacon, Lidar,
  and Whaley}}]{kempe:01}
\bibinfo{author}{\bibfnamefont{J.}~\bibnamefont{Kempe}},
  \bibinfo{author}{\bibfnamefont{D.}~\bibnamefont{Bacon}},
  \bibinfo{author}{\bibfnamefont{D.~A.} \bibnamefont{Lidar}}, \bibnamefont{and}
  \bibinfo{author}{\bibfnamefont{K.~B.} \bibnamefont{Whaley}},
  \bibinfo{journal}{Phys. Rev. A} \textbf{\bibinfo{volume}{63}},
  \bibinfo{pages}{042307} (\bibinfo{year}{2001}).

\bibitem[{\citenamefont{Knill et~al.}(2000)\citenamefont{Knill, Laflamme, and
  Viola}}]{knill:00}
\bibinfo{author}{\bibfnamefont{E.}~\bibnamefont{Knill}},
  \bibinfo{author}{\bibfnamefont{R.}~\bibnamefont{Laflamme}}, \bibnamefont{and}
  \bibinfo{author}{\bibfnamefont{L.}~\bibnamefont{Viola}},
  \bibinfo{journal}{Phys. Rev. Lett.} \textbf{\bibinfo{volume}{84}},
  \bibinfo{pages}{2525} (\bibinfo{year}{2000}).

\bibitem[{\citenamefont{Shabani and Lidar}(2005)}]{shabani:05}
\bibinfo{author}{\bibfnamefont{A.}~\bibnamefont{Shabani}} \bibnamefont{and}
  \bibinfo{author}{\bibfnamefont{D.~A.} \bibnamefont{Lidar}},
  \bibinfo{journal}{Phys. Rev. A} \textbf{\bibinfo{volume}{72}},
  \bibinfo{pages}{042303} (\bibinfo{year}{2005}).

\bibitem[{\citenamefont{Kwiat et~al.}(2000)\citenamefont{Kwiat, Berglund,
  Altepeter, and White}}]{kwiat:00}
\bibinfo{author}{\bibfnamefont{P.~G.} \bibnamefont{Kwiat}},
  \bibinfo{author}{\bibfnamefont{A.~J.} \bibnamefont{Berglund}},
  \bibinfo{author}{\bibfnamefont{J.~B.} \bibnamefont{Altepeter}},
  \bibnamefont{and} \bibinfo{author}{\bibfnamefont{A.~G.} \bibnamefont{White}},
  \bibinfo{journal}{Science} \textbf{\bibinfo{volume}{290}},
  \bibinfo{pages}{498} (\bibinfo{year}{2000}).

\bibitem[{\citenamefont{Zhang et~al.}(2006)\citenamefont{Zhang, Yin, Chen, Lu,
  Zhang, Li, Yang, Wang, and Pan}}]{zhang:06}
\bibinfo{author}{\bibfnamefont{Q.}~\bibnamefont{Zhang}},
  \bibinfo{author}{\bibfnamefont{J.}~\bibnamefont{Yin}},
  \bibinfo{author}{\bibfnamefont{T.-Y.} \bibnamefont{Chen}},
  \bibinfo{author}{\bibfnamefont{S.}~\bibnamefont{Lu}},
  \bibinfo{author}{\bibfnamefont{J.}~\bibnamefont{Zhang}},
  \bibinfo{author}{\bibfnamefont{X.-Q.} \bibnamefont{Li}},
  \bibinfo{author}{\bibfnamefont{T.}~\bibnamefont{Yang}},
  \bibinfo{author}{\bibfnamefont{X.-B.} \bibnamefont{Wang}}, \bibnamefont{and}
  \bibinfo{author}{\bibfnamefont{J.-W.} \bibnamefont{Pan}},
  \bibinfo{journal}{Phys. Rev. A} \textbf{\bibinfo{volume}{73}},
  \bibinfo{pages}{020301(R)} (\bibinfo{year}{2006}).

\bibitem[{\citenamefont{Altepeter et~al.}(2004)\citenamefont{Altepeter, Hadley,
  Wendelken, Berglund, and Kwiat}}]{altepeter:04}
\bibinfo{author}{\bibfnamefont{J.~B.} \bibnamefont{Altepeter}},
  \bibinfo{author}{\bibfnamefont{P.~G.} \bibnamefont{Hadley}},
  \bibinfo{author}{\bibfnamefont{S.~M.} \bibnamefont{Wendelken}},
  \bibinfo{author}{\bibfnamefont{A.~J.} \bibnamefont{Berglund}},
  \bibnamefont{and} \bibinfo{author}{\bibfnamefont{P.~G.} \bibnamefont{Kwiat}},
  \bibinfo{journal}{Phys. Rev. Lett.} \textbf{\bibinfo{volume}{92}},
  \bibinfo{pages}{147901} (\bibinfo{year}{2004}).

\bibitem[{\citenamefont{Mohseni et~al.}(2003)\citenamefont{Mohseni, Lundeen,
  Resch, and Steinberg}}]{mohseni:03}
\bibinfo{author}{\bibfnamefont{M.}~\bibnamefont{Mohseni}},
  \bibinfo{author}{\bibfnamefont{J.~S.} \bibnamefont{Lundeen}},
  \bibinfo{author}{\bibfnamefont{K.~J.} \bibnamefont{Resch}}, \bibnamefont{and}
  \bibinfo{author}{\bibfnamefont{A.~M.} \bibnamefont{Steinberg}},
  \bibinfo{journal}{Phys. Rev. Lett.} \textbf{\bibinfo{volume}{91}},
  \bibinfo{pages}{187903} (\bibinfo{year}{2003}).

\bibitem[{\citenamefont{Viola et~al.}(2001)\citenamefont{Viola, Fortunato,
  Pravia, Knill, Laflamme, and Cory}}]{viola:01}
\bibinfo{author}{\bibfnamefont{L.}~\bibnamefont{Viola}},
  \bibinfo{author}{\bibfnamefont{E.~M.} \bibnamefont{Fortunato}},
  \bibinfo{author}{\bibfnamefont{M.~A.} \bibnamefont{Pravia}},
  \bibinfo{author}{\bibfnamefont{E.}~\bibnamefont{Knill}},
  \bibinfo{author}{\bibfnamefont{R.}~\bibnamefont{Laflamme}}, \bibnamefont{and}
  \bibinfo{author}{\bibfnamefont{D.~G.} \bibnamefont{Cory}},
  \bibinfo{journal}{Science} \textbf{\bibinfo{volume}{293}},
  \bibinfo{pages}{2059} (\bibinfo{year}{2001}).

\bibitem[{\citenamefont{Wei et~al.}(2005)\citenamefont{Wei, Luo, Sun, Zeng,
  Zhan, and Liu}}]{wei:05}
\bibinfo{author}{\bibfnamefont{D.}~\bibnamefont{Wei}},
  \bibinfo{author}{\bibfnamefont{J.}~\bibnamefont{Luo}},
  \bibinfo{author}{\bibfnamefont{X.}~\bibnamefont{Sun}},
  \bibinfo{author}{\bibfnamefont{X.}~\bibnamefont{Zeng}},
  \bibinfo{author}{\bibfnamefont{M.}~\bibnamefont{Zhan}}, \bibnamefont{and}
  \bibinfo{author}{\bibfnamefont{M.}~\bibnamefont{Liu}},
  \bibinfo{journal}{Phys. Rev. Lett.} \textbf{\bibinfo{volume}{95}},
  \bibinfo{pages}{020501} (\bibinfo{year}{2005}).

\bibitem[{\citenamefont{Ollerenshaw et~al.}(2003)\citenamefont{Ollerenshaw,
  Lidar, and Kay}}]{ollerenshaw:03}
\bibinfo{author}{\bibfnamefont{J.~E.} \bibnamefont{Ollerenshaw}},
  \bibinfo{author}{\bibfnamefont{D.~A.} \bibnamefont{Lidar}}, \bibnamefont{and}
  \bibinfo{author}{\bibfnamefont{L.~E.} \bibnamefont{Kay}},
  \bibinfo{journal}{Phys. Rev. Lett.} \textbf{\bibinfo{volume}{91}},
  \bibinfo{pages}{217904} (\bibinfo{year}{2003}).

\bibitem[{\citenamefont{Kielpinski et~al.}(2001)\citenamefont{Kielpinski,
  Meyer, Rowe, Sackett, Itano, Monroe, and Wineland}}]{kielpinski:01}
\bibinfo{author}{\bibfnamefont{D.}~\bibnamefont{Kielpinski}},
  \bibinfo{author}{\bibfnamefont{V.}~\bibnamefont{Meyer}},
  \bibinfo{author}{\bibfnamefont{M.~A.} \bibnamefont{Rowe}},
  \bibinfo{author}{\bibfnamefont{C.~A.} \bibnamefont{Sackett}},
  \bibinfo{author}{\bibfnamefont{W.~M.} \bibnamefont{Itano}},
  \bibinfo{author}{\bibfnamefont{C.}~\bibnamefont{Monroe}}, \bibnamefont{and}
  \bibinfo{author}{\bibfnamefont{D.~J.} \bibnamefont{Wineland}},
  \bibinfo{journal}{Science} \textbf{\bibinfo{volume}{291}},
  \bibinfo{pages}{1013} (\bibinfo{year}{2001}).

\bibitem[{\citenamefont{Langer et~al.}(2005)\citenamefont{Langer, Ozeri, Jost,
  Chiaverini, DeMarco, Ben-Kish, Blakestad, Britton, Hume, Itano
  et~al.}}]{langer:05}
\bibinfo{author}{\bibfnamefont{C.}~\bibnamefont{Langer}},
  \bibinfo{author}{\bibfnamefont{R.}~\bibnamefont{Ozeri}},
  \bibinfo{author}{\bibfnamefont{J.~D.} \bibnamefont{Jost}},
  \bibinfo{author}{\bibfnamefont{J.}~\bibnamefont{Chiaverini}},
  \bibinfo{author}{\bibfnamefont{B.}~\bibnamefont{DeMarco}},
  \bibinfo{author}{\bibfnamefont{A.}~\bibnamefont{Ben-Kish}},
  \bibinfo{author}{\bibfnamefont{R.~B.} \bibnamefont{Blakestad}},
  \bibinfo{author}{\bibfnamefont{J.}~\bibnamefont{Britton}},
  \bibinfo{author}{\bibfnamefont{D.~B.} \bibnamefont{Hume}},
  \bibinfo{author}{\bibfnamefont{W.~M.} \bibnamefont{Itano}},
  \bibnamefont{et~al.}, \bibinfo{journal}{Phys. Rev. Lett.}
  \textbf{\bibinfo{volume}{95}}, \bibinfo{pages}{060502}
  (\bibinfo{year}{2005}).

\bibitem[{\citenamefont{Bargatin et~al.}(2000)\citenamefont{Bargatin,
  Grishanin, and Zadkov}}]{bargatin:00}
\bibinfo{author}{\bibfnamefont{I.~V.} \bibnamefont{Bargatin}},
  \bibinfo{author}{\bibfnamefont{B.~A.} \bibnamefont{Grishanin}},
  \bibnamefont{and} \bibinfo{author}{\bibfnamefont{V.~N.}
  \bibnamefont{Zadkov}}, \bibinfo{journal}{Phys. Rev. A}
  \textbf{\bibinfo{volume}{61}}, \bibinfo{pages}{052305}
  (\bibinfo{year}{2000}).

\bibitem[{\citenamefont{Ficek and Tana\'{s}}(2002)}]{ficek:02}
\bibinfo{author}{\bibfnamefont{Z.}~\bibnamefont{Ficek}} \bibnamefont{and}
  \bibinfo{author}{\bibfnamefont{R.}~\bibnamefont{Tana\'{s}}},
  \bibinfo{journal}{Phys. Rep.} \textbf{\bibinfo{volume}{372}},
  \bibinfo{pages}{369} (\bibinfo{year}{2002}).

\bibitem[{\citenamefont{Lukin and Hemmer}(2000)}]{lukin:00}
\bibinfo{author}{\bibfnamefont{M.~D.} \bibnamefont{Lukin}} \bibnamefont{and}
  \bibinfo{author}{\bibfnamefont{P.~R.} \bibnamefont{Hemmer}},
  \bibinfo{journal}{Phys. Rev. Lett.} \textbf{\bibinfo{volume}{84}},
  \bibinfo{pages}{2818} (\bibinfo{year}{2000}).

\bibitem[{\citenamefont{Beige et~al.}(2000)\citenamefont{Beige, Huelga, Knight,
  Plenio, and Thompson}}]{beige:00}
\bibinfo{author}{\bibfnamefont{A.}~\bibnamefont{Beige}},
  \bibinfo{author}{\bibfnamefont{S.~F.} \bibnamefont{Huelga}},
  \bibinfo{author}{\bibfnamefont{P.~L.} \bibnamefont{Knight}},
  \bibinfo{author}{\bibfnamefont{M.~B.} \bibnamefont{Plenio}},
  \bibnamefont{and} \bibinfo{author}{\bibfnamefont{R.~C.}
  \bibnamefont{Thompson}}, \bibinfo{journal}{J. Mod. Opt.}
  \textbf{\bibinfo{volume}{47}}, \bibinfo{pages}{401} (\bibinfo{year}{2000}).

\bibitem[{\citenamefont{Brennen et~al.}(1999)\citenamefont{Brennen, Caves,
  Jessen, and Deutsch}}]{brennen:99}
\bibinfo{author}{\bibfnamefont{G.~K.} \bibnamefont{Brennen}},
  \bibinfo{author}{\bibfnamefont{C.~M.} \bibnamefont{Caves}},
  \bibinfo{author}{\bibfnamefont{P.~S.} \bibnamefont{Jessen}},
  \bibnamefont{and} \bibinfo{author}{\bibfnamefont{I.~H.}
  \bibnamefont{Deutsch}}, \bibinfo{journal}{Phys. Rev. Lett.}
  \textbf{\bibinfo{volume}{82}}, \bibinfo{pages}{1060} (\bibinfo{year}{1999}).

\bibitem[{\citenamefont{Jaksch et~al.}(2000)\citenamefont{Jaksch, Cirac,
  Zoller, Rolston, C\^ot\'e, and Lukin}}]{jaksch:00}
\bibinfo{author}{\bibfnamefont{D.}~\bibnamefont{Jaksch}},
  \bibinfo{author}{\bibfnamefont{J.~I.} \bibnamefont{Cirac}},
  \bibinfo{author}{\bibfnamefont{P.}~\bibnamefont{Zoller}},
  \bibinfo{author}{\bibfnamefont{S.~L.} \bibnamefont{Rolston}},
  \bibinfo{author}{\bibfnamefont{R.}~\bibnamefont{C\^ot\'e}}, \bibnamefont{and}
  \bibinfo{author}{\bibfnamefont{M.~D.} \bibnamefont{Lukin}},
  \bibinfo{journal}{Phys. Rev. Lett.} \textbf{\bibinfo{volume}{85}},
  \bibinfo{pages}{2208} (\bibinfo{year}{2000}).

\bibitem[{\citenamefont{Barenco et~al.}(1995)\citenamefont{Barenco, Deutsch,
  Ekert, and Jozsa}}]{barenco:95}
\bibinfo{author}{\bibfnamefont{A.}~\bibnamefont{Barenco}},
  \bibinfo{author}{\bibfnamefont{D.}~\bibnamefont{Deutsch}},
  \bibinfo{author}{\bibfnamefont{A.}~\bibnamefont{Ekert}}, \bibnamefont{and}
  \bibinfo{author}{\bibfnamefont{R.}~\bibnamefont{Jozsa}},
  \bibinfo{journal}{Phys. Rev. Lett.} \textbf{\bibinfo{volume}{74}},
  \bibinfo{pages}{4083} (\bibinfo{year}{1995}).

\bibitem[{aga()}]{agarwal:qst2}
\bibinfo{note}{G. S. Agarwal, in \textit{Quantum Statistical Theories of
  Spontaneous Emission and Their Relation to Other Approaches}, edited by G.
  H\"ohler (Springer, Berlin, 1974).}

\bibitem[{\citenamefont{Ficek and Swain}(2005)}]{ficek:int}
\bibinfo{author}{\bibfnamefont{Z.}~\bibnamefont{Ficek}} \bibnamefont{and}
  \bibinfo{author}{\bibfnamefont{S.}~\bibnamefont{Swain}},
  \emph{\bibinfo{title}{Quantum Interference and Coherence}}
  (\bibinfo{publisher}{Springer}, \bibinfo{address}{New York},
  \bibinfo{year}{2005}).

\bibitem[{\citenamefont{Mandel and Wolf}(1995)}]{mandel:qo}
\bibinfo{author}{\bibfnamefont{L.}~\bibnamefont{Mandel}} \bibnamefont{and}
  \bibinfo{author}{\bibfnamefont{E.}~\bibnamefont{Wolf}},
  \emph{\bibinfo{title}{Optical coherence and quantum optics}}
  (\bibinfo{publisher}{Cambridge University Press}, \bibinfo{address}{London},
  \bibinfo{year}{1995}).

\bibitem[{\citenamefont{Dicke}(1954)}]{dicke:54}
\bibinfo{author}{\bibfnamefont{R.~H.} \bibnamefont{Dicke}},
  \bibinfo{journal}{Phys. Rev.} \textbf{\bibinfo{volume}{93}},
  \bibinfo{pages}{99} (\bibinfo{year}{1954}).

\bibitem[{\citenamefont{Zanardi}(1997)}]{zanardi:97b}
\bibinfo{author}{\bibfnamefont{P.}~\bibnamefont{Zanardi}},
  \bibinfo{journal}{Phys. Rev. A} \textbf{\bibinfo{volume}{56}},
  \bibinfo{pages}{4445} (\bibinfo{year}{1997}).

\bibitem[{\citenamefont{Duan and Guo}(1998)}]{duan:98b}
\bibinfo{author}{\bibfnamefont{L.-M.} \bibnamefont{Duan}} \bibnamefont{and}
  \bibinfo{author}{\bibfnamefont{G.-C.} \bibnamefont{Guo}},
  \bibinfo{journal}{Phys. Rev. A} \textbf{\bibinfo{volume}{58}},
  \bibinfo{pages}{3491} (\bibinfo{year}{1998}).

\bibitem[{kif()}]{kiffner:06}
\bibinfo{note}{M. Kiffner, J. Evers, and C. H. Keitel, arXiv:quant-ph/0611071}.

\bibitem[{\citenamefont{DeVoe and Brewer}(1996)}]{devoe:96}
\bibinfo{author}{\bibfnamefont{R.~G.} \bibnamefont{DeVoe}} \bibnamefont{and}
  \bibinfo{author}{\bibfnamefont{R.~G.} \bibnamefont{Brewer}},
  \bibinfo{journal}{Phys. Rev. Lett.} \textbf{\bibinfo{volume}{76}},
  \bibinfo{pages}{2049} (\bibinfo{year}{1996}).

\bibitem[{\citenamefont{Hettich et~al.}(2002)\citenamefont{Hettich, Schmitt,
  Zitzmann, K\"uhn, Gerhardt, and Sandoghdar}}]{hettich:02}
\bibinfo{author}{\bibfnamefont{C.}~\bibnamefont{Hettich}},
  \bibinfo{author}{\bibfnamefont{C.}~\bibnamefont{Schmitt}},
  \bibinfo{author}{\bibfnamefont{J.}~\bibnamefont{Zitzmann}},
  \bibinfo{author}{\bibfnamefont{S.}~\bibnamefont{K\"uhn}},
  \bibinfo{author}{\bibfnamefont{I.}~\bibnamefont{Gerhardt}}, \bibnamefont{and}
  \bibinfo{author}{\bibfnamefont{V.}~\bibnamefont{Sandoghdar}},
  \bibinfo{journal}{Nature} \textbf{\bibinfo{volume}{298}},
  \bibinfo{pages}{385} (\bibinfo{year}{2002}).

\bibitem[{\citenamefont{Eschner et~al.}(2001)\citenamefont{Eschner, Raab,
  Schmidt-Kaler, and Blatt}}]{eschner:01}
\bibinfo{author}{\bibfnamefont{J.}~\bibnamefont{Eschner}},
  \bibinfo{author}{\bibfnamefont{C.}~\bibnamefont{Raab}},
  \bibinfo{author}{\bibfnamefont{F.}~\bibnamefont{Schmidt-Kaler}},
  \bibnamefont{and} \bibinfo{author}{\bibfnamefont{R.}~\bibnamefont{Blatt}},
  \bibinfo{journal}{Nature} \textbf{\bibinfo{volume}{413}},
  \bibinfo{pages}{495} (\bibinfo{year}{2001}).

\bibitem[{\citenamefont{Sakurai}(1994)}]{sakurai:mqm}
\bibinfo{author}{\bibfnamefont{J.~J.} \bibnamefont{Sakurai}},
  \emph{\bibinfo{title}{Modern Quantum Mechanics}}
  (\bibinfo{publisher}{Addison-Wesley}, \bibinfo{address}{Reading, MA},
  \bibinfo{year}{1994}).

\bibitem[{\citenamefont{Evers et~al.}(2006)\citenamefont{Evers, Kiffner,
  Macovei, and Keitel}}]{evers:06}
\bibinfo{author}{\bibfnamefont{J.}~\bibnamefont{Evers}},
  \bibinfo{author}{\bibfnamefont{M.}~\bibnamefont{Kiffner}},
  \bibinfo{author}{\bibfnamefont{M.}~\bibnamefont{Macovei}}, \bibnamefont{and}
  \bibinfo{author}{\bibfnamefont{C.~H.} \bibnamefont{Keitel}},
  \bibinfo{journal}{Phys. Rev. A} \textbf{\bibinfo{volume}{73}},
  \bibinfo{pages}{023804} (\bibinfo{year}{2006}).

\bibitem[{\citenamefont{Agarwal and Patnaik}(2001)}]{agarwal:01}
\bibinfo{author}{\bibfnamefont{G.~S.} \bibnamefont{Agarwal}} \bibnamefont{and}
  \bibinfo{author}{\bibfnamefont{A.~K.} \bibnamefont{Patnaik}},
  \bibinfo{journal}{Phys. Rev. A} \textbf{\bibinfo{volume}{63}},
  \bibinfo{pages}{043805} (\bibinfo{year}{2001}).

\bibitem[{\citenamefont{Akram et~al.}(2000)\citenamefont{Akram, Ficek, and
  Swain}}]{akram:00}
\bibinfo{author}{\bibfnamefont{U.}~\bibnamefont{Akram}},
  \bibinfo{author}{\bibfnamefont{Z.}~\bibnamefont{Ficek}}, \bibnamefont{and}
  \bibinfo{author}{\bibfnamefont{S.}~\bibnamefont{Swain}},
  \bibinfo{journal}{Phys. Rev. A} \textbf{\bibinfo{volume}{62}},
  \bibinfo{pages}{013413} (\bibinfo{year}{2000}).

\bibitem[{\citenamefont{Wootters}(1998)}]{wooters:98}
\bibinfo{author}{\bibfnamefont{W.~K.} \bibnamefont{Wootters}},
  \bibinfo{journal}{Phys. Rev. Lett.} \textbf{\bibinfo{volume}{80}},
  \bibinfo{pages}{2245} (\bibinfo{year}{1998}).

\bibitem[{\citenamefont{Rungta et~al.}(2001)\citenamefont{Rungta, Bu\v{z}ek,
  Caves, Hillery, and Milburn}}]{rungta:01}
\bibinfo{author}{\bibfnamefont{P.}~\bibnamefont{Rungta}},
  \bibinfo{author}{\bibfnamefont{V.}~\bibnamefont{Bu\v{z}ek}},
  \bibinfo{author}{\bibfnamefont{C.~M.} \bibnamefont{Caves}},
  \bibinfo{author}{\bibfnamefont{M.}~\bibnamefont{Hillery}}, \bibnamefont{and}
  \bibinfo{author}{\bibfnamefont{G.~J.} \bibnamefont{Milburn}},
  \bibinfo{journal}{Phys. Rev. A} \textbf{\bibinfo{volume}{64}},
  \bibinfo{pages}{042315} (\bibinfo{year}{2001}).

\end{thebibliography}
\end{document}